\documentclass[10pt,a4paper,twoside]{article}        %
\makeatletter
\usepackage{amsmath}%
\usepackage{amsopn}%
\usepackage{amssymb}
\usepackage{amstext}%
\usepackage{calc}%
\usepackage{dsfont}%
\usepackage{epsfig,epic}
\usepackage{fancyhdr}
\usepackage{txfonts}    %
\usepackage{ntheorem}
\usepackage[latin1]{inputenc}
\usepackage[T1]{fontenc}
\usepackage[english]{babel}
\newlength{\rilegatura}
\newlength{\simmetricspaceh}
\DeclareMathOperator*{\convhull}{conv\ hull}
\newcommand{\supop}[2]{\genfrac{}{}{0pt}{}{\scriptstyle #1}{\scriptstyle #2}}
\newcommand{\newatop}[2]{\genfrac{}{}{0pt}{}{#1}{#2}}   %
\newcommand{\pt}[1]{\left( #1 \right)}                   %
\newcommand{\pq}[1]{\left[ #1 \right]}                   %
\newcommand{\pg}[1]{\left\{ #1 \right\}}                 %
\newcommand{\bs}[1]{\boldsymbol{#1}}
\newcommand{\ket}[1]{\left|\left. #1 \right.\right\rangle}        %
\newcommand{\bra}[1]{\left\langle\left. #1 \right.\right|}        %
\newcommand{\ud}{\mathrm{d}}

\newcommand{\Tr}{\mathrm{Tr}}

\newcommand{\co}[1]{\textsf{#1}}
\newcommand{\qqeedd}{\rule[0pt]{7pt}{7pt}}
\newcommand{\coleq}{\coloneqq}       %
\newcommand{\eqcol}{\eqqcolon}       %
\renewcommand{\geq}{\geqslant}       %
\renewcommand{\leq}{\leqslant}       %
{\theoremstyle{break} \theorembodyfont{\rmfamily}
\newtheorem{DDD}{Definition}
}
{\theoremstyle{break} \theorembodyfont{\rmfamily}
\newtheorem*{NNN}{Remark}
}
{\theoremstyle{break} \theorembodyfont{\rmfamily}
\newtheorem{LLL}{Lemma}}
{\theoremstyle{break} \theorembodyfont{\rmfamily}
\newtheorem{PRS}{Properties}}
{\theoremstyle{break} \theorembodyfont{\rmfamily}
\newtheorem*{PPP}{Proposition}
}
{\theoremstyle{plain}
\newtheorem*{TT}{Theorem}}
\newenvironment{Ventry}[1]%
    {\begin{list}{}{%
        \settowidth{\labelwidth}{#1}%
        \setlength{\leftmargin}{\labelwidth}}}%
    {\end{list}}%
\makeatother
\begin{document}

\lhead[\fancyplain{}{\thepage \protect\hspace{5mm}}]%
      {\fancyplain{}{}}
\rhead[\fancyplain{}{}]%
      {\fancyplain{}{\thepage}}
\chead{}\lfoot{}\cfoot{}\rfoot{}

\

\vspace{-3cm}

\begin{center}
{\bfseries \large Subnormalized states and trace--nonincreasing
maps}\\[4ex]

 {\normalsize Valerio Cappellini$^{1,2}$,
    Hans-J{\"u}rgen Sommers$^3$ and Karol {\.Z}yczkowski$^{1,4}$\\
 {\small\itshape $^1$Centrum Fizyki Teoretycznej, Polska Akademia
 Nauk,  Al. Lotnik{\'o}w 32/44, 02-668 Warszawa, Poland}\\
 {\small\itshape $^2$``Mark Kac'' Complex Systems Research Centre, Uniwersytet
 Jagiello{\'n}ski, ul. Reymonta 4, 30-059 Kraków, Poland}\\
 {\small\itshape $^3$Fachbereich Physik,
 Universit\"{a}t Duisburg-Essen, Campus Duisburg,
   47048 Duisburg, Germany}\\
 {\small\itshape $^4$Instytut Fizyki im. Smoluchowskiego,
 Uniwersytet Jagiello{\'n}ski,
 ul. Reymonta 4, 30-059 Krak{\'o}w, Poland}\\
{\small (Dated: April 02, 2007)}}

\begin{minipage}[c]{14.3cm}
\ \\[1ex]
\small \phantom{We}We investigate the set of completely positive,
trace--nonincreasing linear maps acting on the set ${\cal M}_N$ of
mixed quantum states of size $N$. Extremal point of this set of maps
are characterized and its volume with respect to the
\co{H}ilbert--\co{S}chmidt (Euclidean) measure is computed
explicitly for an arbitrary $N$. The spectra of partially reduced
rescaled dynamical matrices associated with trace--nonincreasing
completely positive maps belong to the $N$--cube inscribed in the
set of subnormalized states of size $N$.
 As a by--product we derive the
measure in ${\cal M}_N$ induced by partial trace of mixed quantum
states distributed uniformly with respect to \co{HS}--measure in
 ${\cal M}_{N^2}$.
\end{minipage}
\end{center}
\thispagestyle{empty}\ \\[-3ex]
\begin{center}
{\small e-mail: valerio@cft.edu.pl \ \ \
h.j.sommers@uni-due.de
  \ \  \ karol@cft.edu.pl}
\end{center}
\section{Introduction}%
Modern application of quantum mechanics increased interest in the
space of quantum states: positive operators normalized by the
assumption that their trace is fixed, $\Tr\:\rho=1$. For
applications in the theory of quantum information processing it is
often sufficient to restrict the attention to the operators acting
on a finite dimensional Hilbert space.

The set ${\cal M}_N$ of density matrices of size $N$ forms a  convex
body embedded in $\mathds{R}^{N^2-1}$. In other words it forms a
cross-section of the cone of positive operators with a hyperplane
corresponding to the normalization condition. In the simplest case
of one qubit the set ${\cal M}_2$  is equivalent, with respect to
the Hilbert-Schmidt (Euclidean) geometry, to a three dimensional
ball, $\mathbf{B}_3$. For higher dimensions the geometry of ${\cal
M}_N$ gets more complicated and differs from the ball
$\mathbf{B}_{N^2-1}$~\cite{GMK05,BZ06}.

Non trivial properties of the set of mixed quantum states attracted
recently a lot of attention. The volume $V$, the hyperarea $A$ and
the radius $R$ of the maximal ball inscribed into ${\cal M}_N$ was
computed with respect to the \co{H}ilbert-\co{S}chmidt
measure~\cite{ZS03}, which leads to the Euclidean geometry. The
volume of the set of quantum states was computed with respect to the
\co{B}ures measure related to quantum
distinguishability~\cite{SZ03}, and also with respect to a wide
class of measures induced by monotone Riemanian
metrics~\cite{Ca01,An06}. The set ${\cal M}_N$ is known to be of a
constant height~\cite{SBZ06}, so the ratio $A/V$ coincides with the
dimensionality of this set, equal to $N^2-1$.

If $N$ is a composite number, the density operators from ${\cal
M}_N$ can represent states of a composed physical system. In this
case one defines the set of separable states, which forms a convex
subset of the set of all states, ${\cal M}_N^{\text{\co{sep}}}
\subset {\cal M}_N^{\phantom{^{\text{\co{p}}}}}$\ . Although a lot
of work has been done to estimate the volume of the subset of
separable states~\cite{ZHSL98, Sl99,Zy99, GB02, Sza04,AS05}, the
problem of finding the exact value of the ratio
 $\mathrm{Vol}\pt{{\cal M}_N^{\text{\co{sep}}}}
/ \mathrm{Vol}\pt{{\cal M}_N^{\phantom{^{\text{\co{p}}}}}}$
remains open even in the simplest case of two qubits~\cite{Sl05}.

In parallel with investigation of the set of quantum states, one
studies properties of the set of completely positive  (\co{CP}) maps
which act on ${\cal M}_N$. Such maps are important not only from the
theoretical point of view:  for instance linear \co{CP} maps acting
on a two--level quantum system correspond to linear optical devices
used in polarisation optics \cite{APW04}.
 Due to Jamio{\l}kowski
isomorphism~\cite{Ja72,ZB04}, the set of trace preserving \co{CP}
maps acting on ${\cal M}_N$ forms a $N^4-N^2$ dimensional subset of
the $N^4-1$ dimensional set ${\cal M}_{N^2}$ of states acting on an
extended Hilbert space, ${\cal H}_N \otimes {\cal H}_N$. In the
simplest case of $N=2$ the structure of this $12$--dimensional
convex set of maps was studied in~\cite{RSW02}.

We start this paper by reviewing the properties of the set of
subnormalized states for which $\Tr\:\rho \leq 1$. Such states are
obtained by taking the convex hull of the set of normalized states
and the particular ``zero state''. In the classical case, one could
argue that such a step is equivalent to increasing the number of
distinguishable events by one, and renaming $0$ into $N+1$. This
reasoning is based on the fact that the set of subnormalized states
of size $N$ as well as the set of normalized states of size $N+1$
form $N$--dimensional simplices. However, this is not the case in
the quantum set--up, in which the set of subnormalized states ${\cal
M}_N^{\text{\co{sub}}}$ has $N^2$ dimensions, in contrast to
$N^2+2N$ dimensional set ${\cal M}_{N+1}$. The fact that the
dimensionality of the set of subnormalized states grows with the
number $N$ of distinguishable states exactly as $N^2$ plays a key
role in an axiomatic approach to quantum mechanics of
Hardy~\cite{Ha01}.

The main aim of this work is to describe the set of completely
positive trace non-increasing maps which act on the set ${\cal M}_N$
of mixed states. We compute the exact volume of this $N^4$
dimensional set with respect to the Euclidean
(\co{H}ilbert--\co{S}chmidt) measure and characterize its extremal
points. The trace non-increasing maps have a realistic physical
motivation, since they describe an experiment, for which with a
certain probability the apparatus does not work. This could be an
interpretation of the ``zero map'' after action of which no result
is recorded. Such maps are sometimes called {\it trace
decreasing}~\cite{NCSB97}, but to emphasize that the set of these
maps contains also all trace preserving maps, we prefer to use a
more precise name of {\it trace non--increasing} maps, (\co{TNI}).

The paper is organized as follows. In Section~\ref{sqs} we analyze
the set of subnormalized states and compute its volume. The measures
in the set of mixed states induced by partial trace are investigated
in Section~\ref{Mibptoms}. In Section~\ref{tnim} we define the set
of trace non--increasing maps and provide its characterization,
while in Section~\ref{Vol_TNIM} the volume of this set is calculated
with respect to the flat (\co{H}ilbert--\co{S}chmidt) measure. The
set of extremal trace non--increasing maps is studied in
Section~\ref{ECPTNIm}.
\section{Subnormalized quantum states}%
\begin{figure}[ht]
\begin{center}
\includegraphics[width=\textwidth]{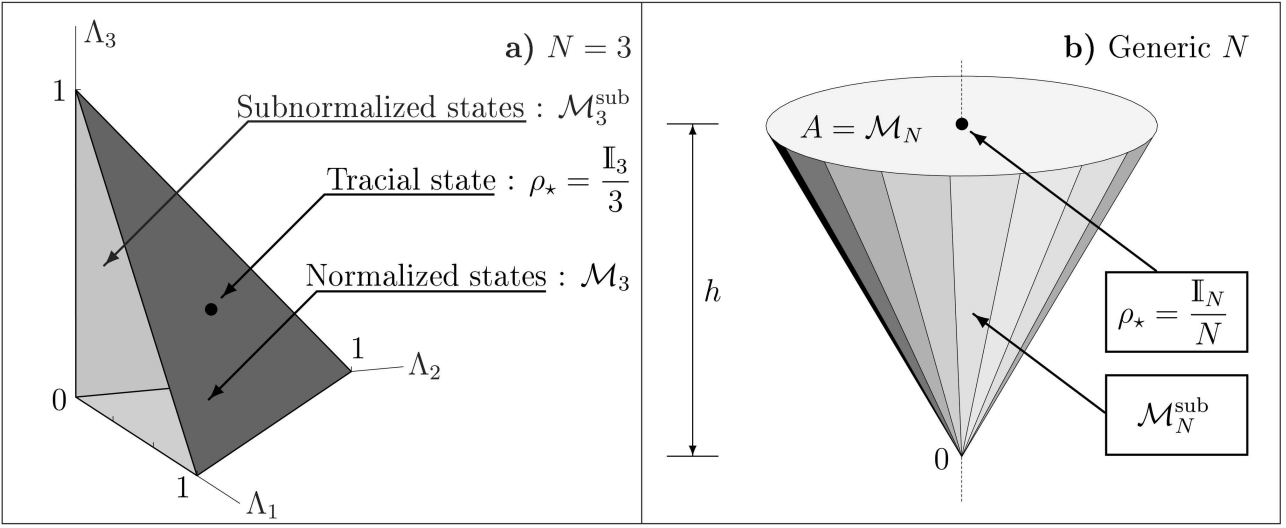}
\caption[]{The set of subnormalized states:
  {\bfseries (a)} the set of eigenvalues for $N=3$,
  {\bfseries (b)} the convex cone of $N^2$ dimensions with zero state at the apex and
  the $\pt{N^2-1}$ dimensional set ${\cal M}_N$ as the base.}%
\label{ch}
\end{center}
\end{figure}

\label{sqs}
Let ${\cal M}_N$ denote the set of normalized quantum states acting
on $N$--dimensional Hilbert space
\begin{equation}
 {\cal M}_N\coleq\big\{\rho\ :\
\rho^{\dagger}\!=\rho,\ \rho\geq 0,\ \Tr\:\rho= 1\big\}
\quad\cdot\nonumber
\end{equation}
${\cal M}_N$ forms a convex set of dimensionality $\pt{N^2-1}$. In
the simplest case $N=2$, this set is equivalent to the \textit{Bloch
ball}, ${\cal M}_2=\mathbf{B}_3\subset\mathds{R}^3$.\\[0ex]
\begin{quote}
\begin{DDD}{}\ \\[-5.5ex]
\begin{Ventry}{}\label{sub_states}
    \item[] An Hermitian, positive operator $\sigma$
  is called a \textit{subnormalized state}, if  $\Tr\: \sigma \leq 1$.
\end{Ventry}
\end{DDD}
\end{quote}
The set of subnormalized states acting on the $N$--dimensional
Hilbert space ${\cal H}_N$ will be denoted by ${\cal
M}_N^{\text{\co{sub}}}$. By construction this set has $N^2$
dimensions and can be defined as a convex hull of the zero operator
and the set of quantum states (see Fig.~\ref{ch}),
\begin{equation}
{\cal M}_N^{\text{\co{sub}}} \coleq\big\{\sigma\ :\
\sigma^{\dagger}\!=\sigma,\ \sigma\geq 0,\
\Tr\:\sigma\leq 1\big\}=\convhull\big\{0, {\cal M}_N\big\} \quad\cdot \label{convhull}%
\end{equation}
For instance, the set  ${\cal M}_2^{\text{\co{sub}}}$ forms a
four--dimensional cone with apex at $0$ and base formed by the Bloch
ball ${\cal M}_2=\mathbf{B}_3$. Note that the dimension of ${\cal
M}_N^{\rm sub}$ grows {\it exactly} as squared dimension of the
Hilbert space. Due to this fact the subnormalized states are a
convenient notion to be used in an axiomatic approach to quantum
theory~\cite{Ha01}.

In order to characterize the set  ${\cal M}_N^{\text{\co{sub}}}$
of subnormalized states we compute in this Section its volume with respect
to the flat \co{HS}--measure
induced by the \co{H}ilbert--\co{S}chmidt metric~\cite{ZS03}.
Consider the set ${\cal M}_2$ of $2\times2$ density matrices,
parameterized by the real {\it Bloch coherence vector} $\vec{\xi}\in
\mathbf{B}_3$,
\begin{equation}
\rho=\frac{\mathds{I}_2}{2} +
\vec{\xi}\cdot\vec{\Xi}\label{Bloch}\quad,
\end{equation}
where $\vec{\Xi}$ denotes the vector of three rescaled traceless
Pauli matrices
$\pt{\sigma_x/\sqrt{2},\sigma_y/\sqrt{2},\sigma_z/\sqrt{2}}$. Note
that with such a normalization the radius of the Bloch ball
$\mathbf{B}_3$ is given by $R_2=1/\sqrt{2}$, as it can be obtained
from the relation $\Tr\:\rho^2\leq1$. With this definition, the
\co{HS}--distance between any two density operators, defined as the
\co{HS} (Frobenius) norm of their difference, proves to be equal to
the Euclidean distance $D_{\text{\co{E}}}(\vec{\xi}_1,\vec{\xi}_2)$
between the labeling Bloch vectors $\vec{\xi}_1,\vec{\xi}_2\in
\mathbf{B}_3\subset\mathds{R}^3$,
\begin{equation}
D_{\text{\co{HS}}}\pt{\rho_{\vec{\xi}_1},\rho_{\vec{\xi}_2}}
=\sqrt{\Tr\pq{{\pt{\rho_{\vec{\xi}_1}-\rho_{\vec{\xi}_2}}}^2}}
=\Big\|{\vec{\xi}_1}-{\vec{\xi}_2}\Big\|
=D_{\text{\co{E}}}(\vec{\xi}_1,\vec{\xi}_2)\quad\cdot\label{hsdist}
\end{equation}
The above formula holds for an arbitrary $N$, provided that
\begin{equation}
\rho=\frac{\mathds{I}_N}{N} +
\vec{\xi}\cdot\vec{\Xi}\label{BlochN}\quad,
\end{equation}
 the real
\textit{coherence vector} $\vec{\xi}$ in~\eqref{BlochN} is taken
$\pt{N^2-1}$--dimensional and $\vec{\Xi}$ now represents an
operator--valued vector which consists of $\pt{N^2-1}$ traceless
Hermitian generators of $\mathrm{SU}\pt{N}$, fulfilling
$\Tr\pt{\Xi_i\Xi_j}=\delta_{ij}$. Note however that in this case the
geometry of the space of coherence vectors $\vec{\xi}$ does not
coincide with the ball $\mathbf{B}_{N^2-1}$, but constitutes instead
a convex subset of it~\cite{BZ06}. The condition $\Tr\:\rho^2\leq1$
yields the upper bound for the length of the coherence vector, $|\xi|\leq\sqrt{(N-1)/N}\eqcol R_N$.
In the case $N=3$ the vector $\vec{\Xi}$ consists of the set of $8$
normalized Gell--Mann matrices ${\pg{\Xi_i}}_{i=1}^8$.

A metric space consisting of a set ${\cal M}_N$ and a distance $d$
is automatically endowed with a measure induced by the metric: The
measure is defined by the assumption that all balls of a fixed
radius defined in ${\cal M}_N$ with respect to the distance $d$ have
the same volume.

The infinitesimal \co{HS}--measure around any matrix $\rho\in{\cal
M}_N$ factorizes as~\cite{ZS03}
\begin{equation}
\ud
\mu_{\text{\co{HS}}}^{\phantom{\text{\co{sub}}}}\pt{\rho}=\frac{\sqrt{N}}{N\,!}\;\ud\nu^{\Delta}\pt{\Lambda_1,\ldots,\Lambda_N}\times\ud\nu^{\text{\co{\,Haar}}}\quad
,\label{vol_el}
\end{equation}
where the factor $\ud\nu^{\Delta}$ represents the measure in the
simplex $\mathbf{\Delta}_{N-1}$ of eigenvalues
$\Lambda_1,\ldots,\Lambda_N$, while $\ud\nu^{\text{\co{\,Haar}}}$
depends on the eigenvectors of $\rho$. The pre--factor $\sqrt{N}$
emerges in~\eqref{vol_el} when we force the variables
$\Lambda_1,\ldots,\Lambda_N$ to live on the simplex
$\mathbf{\Delta}_{N-1}$. This reduces the number of independent
variables by one, since$\sum_{i=1}^N\Lambda_i=1$, and introduces a
factor $\sqrt{\det g}$ in~\eqref{vol_el}, where $g$ denotes the
metric tensor in the $(N-1)$--dimensional simplex. Such a metric
arises due to a change from $N$ linearly dependent variables
${\pg{\Lambda_i}}_{i=1}^{N}$ to the $(N-1)$ linearly independent
ones ${\pg{\Lambda_i}}_{i=1}^{N-1}$.

The last factor $\ud\nu^{\text{\co{\,Haar}}}$ can be integrated on
the entire complex \textit{flag manifold}~\cite{Hu63,ZS03} defined
by the coset space,
$Fl^{(N)}_{\mathds{C}}\coleq\mathrm{U}(N)/{\pq{\mathrm{U}(1)}}^N$,
that is the space of equivalence classes of matrices $U$
diagonalising the given $\rho$.
The volume of the flag manifold induced by the parametrization used
in~\eqref{BlochN} is given by~\cite{ZS03}
\begin{equation}
\mathrm{Vol}\pq{Fl^{(N)}_{\mathds{C}}}=\int_{Fl^{(N)}_{\mathds{C}}}\ud\nu^{\text{\co{\,Haar}}}=
\frac{{(2\pi)}^{N(N-1)/2}}{1\,!\;2\,!\;\cdots(N-1)\,!}\label{flagman}
\quad\cdot
\end{equation}
Even after splitting off the $N$--phases of
${\pq{\mathrm{U}(1)}}^N$, a residual arbitrariness still remains in
the diagonalization of $\rho$, related to the fact that different
permutations of $N$ generically different eigenvalues $\Lambda_i$
belong to the same unitary orbit. This explains the factor $N\,!$
in~\eqref{vol_el}, given by the number of equivalent \textit{Weyl
chambers} of the simplex $\mathbf{\Delta}_{N-1}$.

The measure $\ud\nu^{\Delta}\pt{\Lambda_1,\ldots,\Lambda_N}$
reads~\cite{ZS03}
\begin{equation}
\ud\nu^{\Delta}(\Lambda_1,\ldots,\Lambda_N)=\delta\pt{\sum_{i=1}^N
\Lambda_i -1}\prod_{i=1}^N \:\Theta(\Lambda_i)\prod_{i<j}
\pt{\Lambda_i-\Lambda_j}^2\;\ud\Lambda_1\,\ldots\,\ud\Lambda_N\quad,\label{dimu}
\end{equation}
where the Dirac delta and the product of the Heaviside step
functions $\Theta$ ensure that the measure is concentrated on the
simplex $\mathbf{\Delta}_{N-1}$. Up to a normalization constant, the
r.h.s. of the above equation defines a probability distribution on
the simplex,
\begin{equation}
 \ud\nu^{\Delta}(\Lambda_1,\ldots,\Lambda_N)\propto
P^{(2)}_{\text{\co{HS}}}(\Lambda_1,\ldots,\Lambda_N)\;\ud\Lambda_1\,\ldots\,\ud\Lambda_N\label{prob_hs}
 \quad,
\end{equation}
where the upper index $2$ refers to the exponent in the last factor
of equation~\eqref{dimu}. In random matrix theory it is commonly
denoted by $\beta$ and called \textit{repulsion exponent}, as it
determines the repulsion between adjacent eigenvalues. It is equal
to $1$, $2$ and $4$ for real, respectively complex, respectively
symplectic ensembles of random matrices~\cite{Me91}. In this work we
are going to restrict our attention to complex density matrices so
we fix $\beta=2$, but one may also repeat our analysis for other
universality classes.

For later purpose, we now introduce a bigger family of probability
distributions, indexed by a real parameter $\alpha$:
\begin{subequations}
\label{constab}
\begin{equation}
P^{(\alpha,2)}_{N}(\Lambda_1,\ldots,\Lambda_N) \coleq  \
C_{N}^{(\alpha,2)} \delta\pt{1-\sum_{i=1}^N \Lambda_i}\prod_{i=1}^N
\Theta(\Lambda_i)\:\Lambda_i^{\alpha-1} \prod_{i<j}
\pt{\Lambda_i-\Lambda_j}^2 \quad,%
\end{equation}
with normalization constant $C_N^{(\alpha,2)}$ given by
\begin{equation}
\frac{1}{C_N^{(\alpha,2)} }\coleq \int\delta\pt{1-\sum_{i=1}^N
\Lambda_i}\prod_{i=1}^N \Theta(\Lambda_i)\:\Lambda_i^{\alpha-1}
\prod_{i<j}
\pt{\Lambda_i-\Lambda_j}^2\;\ud\Lambda_1\,\ldots\,\ud\Lambda_N=\frac{\prod_{j=1}^N
\Gamma(1+j) \;\Gamma[j+(\alpha-1)]}{\Gamma[N^2 +(\alpha-1)N]}
     \quad \cdot\label{constab2}
\end{equation}
\end{subequations}
The probability distribution in~\eqref{prob_hs} represents a special
case of $P^{(\alpha,2)}_{N}$, being
\begin{equation}
P^{(2)}_{\text{\co{HS}}}(\Lambda_1,\ldots,\Lambda_N)\ =
\left.P^{(\alpha,2)}_{N}(\Lambda_1,\ldots,\Lambda_N)
\right|_{\alpha=1} \quad .\label{alphau1}
\end{equation}
Putting together equations~(\ref{vol_el}--%
\ref{dimu}) and~\eqref{constab2} one derives~\cite{ZS03} the
(\co{HS}--) volume of the set ${\cal M}_N$ of mixed quantum states
\begin{equation}
\mathrm{Vol}_{\textsf{\:HS}}\pt{{\cal M}_N}=\int_{{\cal
M}_N}\ud\mu_{\text{\co{HS}}}^{\phantom{\text{\co{sub}}}}\pt{\rho}=\frac{\sqrt{N}}{N\,!}\:\frac{\mathrm{Vol}\pq{Fl^{(N)}_{\mathds{C}}}}{C_N^{(1,2)}}=\sqrt{N}\:{(2\pi)}^{N(N-1)/2}
\:\frac{\Gamma\pt{1}\Gamma\pt{2}\cdots\Gamma\pt{N}}{\Gamma\pt{N^2}}\quad\cdot\label{volqs}
\end{equation}
As shown later in this Section, the volume of the set ${\cal
M}_N^{\text{\co{sub}}}$ of subnormalized states can be easily
obtained using the (flat) Euclidean geometry as the volume of the
cone with base given by ${\cal M}_N$\ . However, for later use, we
shall first derive this formula directly by integrating an extension
of the \co{HS}--measure over the entire cone. Let us first formulate
the following
Lemma, proved in Appendix~\ref{app_A1}\\
\begin{quote}
\begin{LLL}{}\ \label{Lemma1}\\
Consider a one parameter family of probability measures
$\ud\nu_K(\Lambda_1,\ldots,\Lambda_N)$ defined on the set
$\text{\co{CH}}\pt{N}\coleq\convhull\big\{0,
\mathbf{\Delta}_{N-1}\big\}$
\begin{equation}
\ud\nu_K(\Lambda_1,\ldots,\Lambda_N)=\prod_{i=1}^N
\:\Theta(\Lambda_i)\: \Lambda_i^{K-N}\prod_{i<j}
\pt{\Lambda_i-\Lambda_j}^2\;\ud\Lambda_1\,\ldots\,\ud\Lambda_N\quad
\label{nu_K}
\end{equation}
and labeled by integers $K\geq N$. Then the volume
$\nu_K\pt{\text{\co{CH}}\pt{N}}$ reads
\begin{equation}
\nu_K\pt{\text{\co{CH}}\pt{N}}=\int_{\text{\co{CH}}\pt{N}}\ud\nu_K(\Lambda_1,\ldots,\Lambda_N) =
\frac{1}{\,KN\:C_N^{(K-N+1,2)}} \nonumber\quad,
\end{equation}
where $C_N^{(K-N+1,2)}$ is the coefficient defined
in~\eqref{constab2}, with $\alpha=K-N+1$.
\end{LLL}
\end{quote}
\noindent The measure $\ud
\mu_{\text{\co{HS}}}^{\text{\co{sub}}}\pt{\sigma}$ on the set of
subnormalized states ${\cal M}_N^{\text{\co{sub}}}$ is given by
\begin{equation}
\ud
\mu_{\text{\co{HS}}}^{\text{\co{sub}}}\pt{\sigma}=\frac{1}{N\,!}\;\ud\nu^{\text{\co{sub}}}
(\Lambda_1,\ldots,\Lambda_N)\times\ud\nu^{\text{\co{\,Haar}}}\quad
,\label{vol_el_sub}
\end{equation}
and differs from the one of equation~\eqref{vol_el}, relative to
quantum normalized states, by the factor $\sqrt{N}$. In
equation~\eqref{vol_el}, such a factor was due to the change of
variables needed to express the volume elements in terms of the set
of $(N-1)$ independent variables on the simplex
$\mathbf{\Delta}_{N-1}$; in the present case we do not need to
change the variables anymore, and the factor $\sqrt{N}$ does not
appear.

Moreover, the definition of the measure $\ud\nu^{\text{\co{sub}}}$
on the entire set $\text{\co{CH}}\pt{N}$ of Lemma~\ref{Lemma1} used
in equation~\eqref{vol_el_sub} differs from the one defined on the
simplex $\mathbf{\Delta}_{N-1}$ and used in equation~\eqref{dimu}.
In particular, for $K=N$ equation~\eqref{nu_K} implies
\begin{equation}
\ud\nu^{\text{\co{sub}}}(\Lambda_1,\ldots,\Lambda_N)=\ud\nu_N(\Lambda_1,\ldots,\Lambda_N)=\Theta\pt{\sum_{i=1}^N
\Lambda_i -1}\prod_{i=1}^N \:\Theta(\Lambda_i)\prod_{i<j}
\pt{\Lambda_i-\Lambda_j}^2\;\ud\Lambda_1\,\ldots\,\ud\Lambda_N\quad\cdot\label{dimuch}
\end{equation}
In analogy to the derivation of the volume~\eqref{volqs} of the set
of normalized states, the \co{HS}--volume of the set ${\cal
M}_N^{\text{\co{sub}}}$ of subnormalized states can be earned by
equations~(\ref{vol_el_sub},~\ref{dimuch},~\ref{flagman})
and~\eqref{constab2} together with Lemma~\ref{Lemma1}, in which we
set $K=N$:
\begin{equation}
\mathrm{Vol}_{\textsf{\:HS}}\pt{{\cal
M}_N^{\text{\co{sub}}}}\ =\ \frac{1}{N\,!}\times
\mathrm{Vol}\pq{Fl^{(N)}_{\mathds{C}}}\times
\nu_N\pt{\text{\co{CH}}\pt{N}}
\ =\ {(2\pi)}^{N(N-1)/2}
\:\frac{\Gamma\pt{1}\Gamma\pt{2}\cdots\Gamma\pt{N}}{N^2\;\Gamma\pt{N^2}}\quad\cdot\label{volsub}
\end{equation}
The above result can be obtained directly using the Archimedean
formula for the Euclidean volume of the $D$--dimensional cone of
Figure~\ref{ch} (panel {\bf b})
\begin{equation}
V=\frac{1}{D}\cdot A\cdot h\quad,\label{volcone}
\end{equation}
where $V$ is the $D$--dimensional volume of the (hyper--) cone
representing ${\cal M}_N^{\text{\co{sub}}}$, $A$ is the area of its
$(D-1)$--dimensional base ${\cal M}_N$, and $h$ denotes its height,
that is the distance between the base and the apex (the latter
corresponding to the state $\sigma_0=0$). Making use of the
definition of \co{HS}--distance~\eqref{hsdist}, one gets the results
\begin{equation}
\begin{cases}
V&=\mathrm{Vol}_{\textsf{\:HS}}\pt{{\cal
M}_N^{\text{\co{sub}}}}\\
D&=\textrm{Dimension}\pt{{\cal M}_N^{\text{\co{sub}}}}=N^2\\
A&=\mathrm{Vol}_{\textsf{\:HS}}\pt{{\cal
M}_N}\\
h&=\displaystyle D_{\text{\co{HS}}}\pt{{\cal
M}_N,0}=\inf_{\rho\in{\cal
M}_N}D_{\text{\co{HS}}}\pt{\rho,0}=\frac{1}{\sqrt{N}}
\end{cases}\quad\label{conequantities}.
\end{equation}
Note that $h= D_{\text{\co{HS}}}\pt{\rho_{\star},0}$, where
$\rho_{\star}=\mathds{I}_N/N$ denotes the maximally mixed state of
Fig.~\ref{ch}, due to the following chain of relations:
\begin{equation}
h^2=\inf_{\rho\in{\cal
M}_N}D_{\text{\co{HS}}}^2\pt{\rho,0}=\inf_{\rho\in {\cal M}_N}
\Tr\:\rho^2=
\inf_{\vec{\Lambda}\in \mathbf{\Delta}_{N-1}}
\sum_{i=1}^{N}\Lambda_i^2=\inf_{\Lambda_1+\cdots+\Lambda_N=1}
\pq{\Lambda_1^2+\cdots\Lambda_N^2}=\frac{1}{N}\quad\cdot
\end{equation}
Substituting results of~\eqref{conequantities} into~\eqref{volcone}
we arrive at
\begin{equation}
\mathrm{Vol}_{\textsf{\:HS}}\pt{{\cal
M}_N^{\text{\co{sub}}}}=\frac{\mathrm{Vol}_{\textsf{\:HS}}\pt{{\cal
M}_N}}{N^{5/2}}\quad\label{volsub2}
\end{equation}
which, due to~\eqref{volqs}, is consistent with the volume given
by~\eqref{volsub}.
\subsection{Generating random subnormalized mixed states with respect to \co{HS} measure} %
\label{GHS}
As we have already seen, the \co{HS}--measure endows ${\cal M}_N$
with the flat, Euclidean geometry. Moreover, as emphasized in
FIG.~\ref{ch}, the set ${\cal M}_N^{\text{\co{sub}}}$ of
subnormalized states constitutes with respect to this geometry an
$N^2$--dimensional cone, whose $\pt{N^2-1}$--dimensional base is
${\cal M}_N$. Thus, directly from~\eqref{convhull}, every state
$\sigma\in{\cal M}_N^{\text{\co{sub}}}$ can be decomposed as
\begin{equation}
\sigma=a\rho+(1-a)\,0\quad,\label{decomp}
\end{equation}
where $\rho\in{\cal M}_N$, $0$ is the null-state and $a$ is a
positive number less or equal to $1$. Therefore generating states
${\pg{\sigma_i}}$ uniformly in the cone means to distribute
homogeneously
 $\rho_i$ in ${\cal M}_N$ and to use them in~\eqref{decomp}.
The weights $a_i\in\pq{0,1}$ must be distributed accordingly to a
density function $f(a)$ scaling as $a^{N^2-1}$. These random numbers
may be obtained by inverting the cumulative distribution function
$F(x)\coleq\int_0^x f(x)\,\ud x$ over uniformly distributed random
numbers $\xi_i\in\pq{0,1}$, that is generating homogeneously
$\xi_i\in\pq{0,1}$ and taking $a_i\coleq
F^{-1}(\xi_i)=\xi_i^{1/N^2}$. As a result, given a sequence of
\co{HS}--distributed mixed states ${\pg{\rho_i}}\in{\cal M}_N$\ , as
the one studied in Section~\ref{PTOPS}, and a sequence
${\pg{\xi_i}}$ of random numbers uniformly distributed in
$\pq{0,1}$, one obtains an algorithmic prescription to generate a
sequence of \co{HS}--distributed random subnormalized states
$\pg{\sigma_i=\xi_i^{1/N^2}\rho_i}\:\cdot$
\section{Measure induced by partial trace of mixed states}%
\label{Mibptoms}
The aim of this Section is to determine the measure induced on a
bipartite $K \times N$ quantum system $AB$, represented by means of
an extended Hilbert space ${\cal H}_{AB}={\cal H}_A\otimes {\cal
H}_B=\mathds{C}^K\otimes \mathds{C}^N$, by partial tracing over one
of the two subsystem. Such measures will be essential in determining
the volumes of various sets of maps, that is done in the
subsequent Sections of this work. Without loss of generality, we
consider ancillary systems  $A$ whose dimension $K$ is greater or
equal to the dimension $N$ of the system $B$. The induced measure
 depends on the choice of the systems to trace over, and on the
initial distribution of the density operators $\rho_{AB}$ on the
composite system ${\cal H}_{AB}$\ . We start by analyzing a
propaedeutic example:
\subsection{Partial tracing over pure states} %
\label{PTOPS} Consider random pure states
$\rho_{AB}=|\psi_{AB}\rangle\langle\psi_{AB}|$ distributed according
to the natural \co{F}ubini--\co{S}tudy (\co{FS}) measure. Such a
measure is the only unitarily invariant one on the space of pure
states, and for $N=2$ it gives the measure uniformly covering the
entire boundary of the Bloch ball~\cite{BZ06}. A general pure state
can be represented in a product basis as
\begin{equation}
 |\psi_{AB}\rangle = \:\sum_{i=1}^K \sum_{j=1}^N \;M_{j\,i}
\;|i\rangle_A \otimes
|j\rangle_B \quad \cdot %
\nonumber%
\end{equation}
The positive matrix $\rho_B=MM^{\dagger}$ is equal to the density
matrix obtained by a partial trace on the $K$--dimensional space
$A$, $\rho_B=\Tr_A \pt{\rho_{AB}}$. The spectrum of $\rho_B$
coincides with the set of Schmidt coefficients $\Lambda_i$ of the
pure state $|\psi_{AB}\rangle$. The matrix $M$ needs not to be
Hermitian, the only constraint is the trace condition,
 $\Tr\: MM^{\dagger}=1$, that makes $\rho_B$ a density matrix. Furthermore,
the natural measure on the space of pure states corresponds to
taking $M$ from the Ginibre ensemble~\cite{Me91} and then
renormalize them to ensure $\Tr\: MM^{\dagger}=1$~\cite{ZS01}. The
probability distribution of the Schmidt coefficients implied by the
\co{FS} measure on ${\cal H}_A\otimes {\cal H}_B$ is given
by~\cite{LP88}
\begin{subequations}
\label{gen_meas}
\begin{equation}
P^{(2)}_{N,K}(\Lambda_1,\ldots,\Lambda_N)\ = \ B^{(2)}_{N,K}\;
\delta\pt{1-\sum_{i} \Lambda_i}
\prod_i\:\Theta(\Lambda_i)\:\Lambda_i^{K-N}\prod_{i<j}{\pt{\Lambda_i-\Lambda_j}}^2
\quad ,%
\end{equation}
in which the upper index 2 is the repulsion exponent, as we deal
with complex density matrices. The normalization constant
$B^{(2)}_{N,K}$ reads~\cite{ZS01}
\begin{equation}
 B^{(2)}_{N,K} \coleq  \frac{\Gamma(KN)}{ \prod_{j=0}^{N-1} \Gamma\pt{K-j}\: \Gamma\pt{N-j+1}} \quad \cdot\label{gen_mes_b}
\end{equation}
\end{subequations}
Observe that the entire
distribution~\eqref{constab}, including the normalization constant, can be tuned
into eq.~\eqref{gen_meas}, provided we choose $\alpha-1=K-N$. In
particular, for $\alpha=1$, that is $K=N$, equation~\eqref{alphau1}
shows that the induced distribution for the eigenvalues of $\rho_B$ coincides with the \co{HS}--distribution. Thus, generating normalized
Wishart matrices $MM^{\dagger}$, with $M$ belonging to the Ginibre
ensemble of $N\times N$ matrices, becomes a useful procedure for
producing $N\times N$ density matrices \co{HS}--distributed, and the
algorithm described in Section~\ref{GHS} becomes effective.

\subsection{Partial tracing over random \co{HS}--distributed mixed states} %
\label{ptorhsdms}
Consider now a related problem of determining the probability distribution of states
$\rho_B=\Tr_A \pt{\rho_{AB}}$, as it is assumed that the mixed states $\rho_{AB}$ are distributed according to the \co{HS}--measure on ${\cal M}_{KN}$.

From Section~\ref{PTOPS} we know that $\rho_{AB}$
itself can be generated by taking \co{FS}--distributed pure state
$\rho_{A^{\prime}B^{\prime}AB}=|\psi_{A^{\prime}B^{\prime}AB}\rangle\langle\psi_{A^{\prime}B^{\prime}AB}|$
in an extended Hilbert space ${\cal
H}_{A^{\prime}B^{\prime}AB}={\cal H}_{A^{\prime}B^{\prime}}\otimes
{\cal H}_{AB}=\mathds{C}^{KN}\otimes \mathds{C}^{KN}$ and by partial
tracing over the ancillary subsystem $A^{\prime}B^{\prime}$. Putting
all together we obtain
\begin{equation}
\rho_B=\Tr_A \pq{\rho_{AB}}=\Tr_{A}\pq{\Tr_{A^{\prime}B^{\prime}}
\pt{\rho_{A^{\prime}B^{\prime}AB}}}=\Tr_{AA^{\prime}B^{\prime}}
\pq{|\psi_{A^{\prime}B^{\prime}AB}\rangle\langle\psi_{A^{\prime}B^{\prime}AB}|}\quad\cdot
\end{equation}
The latter equation implies that the desired distribution is
obtained by coupling the $N$--dimensional system $B$ to
an environment of size $NK^2$, generating \co{FS}--distributed pure
states in this overall $N^2K^2$ system, and finally tracing out the
environment. Hence the distribution of the spectrum of $\rho_B$
is given by $P^{(2)}_{N,NK^2}(\Lambda_1,\ldots,\Lambda_N)$ of
equation~\eqref{gen_meas}.

In the rest of this Section we will consider the special case $K=N$.
This assumption corresponds to generating mixed states distributed according to the \co{HS} measure
on the space ${\cal M}_{N^2}$ of bipartite systems, and tracing out one of them.
In particular we will
focus on the simplest cases of $N=2$ (qubits) and $N=3$
(qutrits).
\subsection{Partial trace of two--qubits mixed quantum states} %
Let us start with the simplest case $N=2$, for which
$B^{(2)}_{N,N^3}=B^{(2)}_{2,8}=180\,180$ and the simplex
$\mathbf{\Delta}_{1}$, corresponding to the positive
$\Lambda_1,\Lambda_2$ such that $\Lambda_1+\Lambda_2=1$, is nothing
but an interval $\pq{0,1}$. The probability distribution~\eqref{gen_meas}
reads in this case
\begin{equation}
P^{(2)}_{2,8}(\Lambda_1,\Lambda_2)\ = \ 180\,180\;
\delta\pt{1-\Lambda_1-\Lambda_2}\;
\Theta(\Lambda_1)\;\Theta(\Lambda_2)\;\Lambda_1^{6}\;\Lambda_2^{6}\;{\pt{\Lambda_1-\Lambda_2}}^2
\quad \cdot\label{p28l}%
\end{equation}
We express the eigenvalues $\Lambda_1$ and $\Lambda_2$ in terms of a
real parameter $r\in\pq{-\frac{1}{2},\frac{1}{2}} \eqcol
\widehat{\mathbf{\Delta}}_1$ as follows
\begin{equation}
\begin{cases}
\Lambda_1=\frac{1}{2}+r\\
\Lambda_2=\frac{1}{2}-r\\
\end{cases}\label{par_r}
\end{equation}
and we earn from~\eqref{p28l} the radial distribution inside the Bloch ball
\begin{equation}
\widetilde{P}(r)\ = \ 720\,720\;\;
\chi_{\widehat{\mathbf{\Delta}}_1}\pt{r}\;{\pt{\frac{1}{4}-r^2}}^6\;r^2
\quad,\label{p28r}%
\end{equation}
where $\chi_{\widehat{\mathbf{\Delta}}_1}\pt{r}$ denotes the
indicator function of the simplex $\widehat{\mathbf{\Delta}}_1$ (see
FIG.~\ref{ch2}a).
\subsection{Partial trace of two--qutrits mixed quantum states} %
For $N=3$ the eigenvalues $\Lambda_1$, $\Lambda_2$ and $\Lambda_3$
can be expressed in polar coordinates $\pt{r,\phi}$ in a form
similar to~\eqref{par_r},
\begin{equation}
\begin{cases}
\Lambda_1=\frac{1}{3}+r\cos\pt{\phi+\frac{2}{3}\pi}\\
\Lambda_2=\frac{1}{3}+r\cos\pt{\phi}\\
\Lambda_3=\frac{1}{3}+r\cos\pt{\phi-\frac{2}{3}\pi}\\
\end{cases}\label{par_r_phi}\quad\cdot
\end{equation}
Similarly as before, we indicate with $\widehat{\mathbf{\Delta}}_2$
the counter--image of the simplex $\mathbf{\Delta}_{2}$, that is the
set in the $\pt{r,\phi}$ plane such that
$\pt{\Lambda_1\pt{r,\phi},\Lambda_2\pt{r,\phi},\Lambda_3\pt{r,\phi}}\in\mathbf{\Delta}_{2}$.

\noindent Computing the Jacobian of the
transformation~\eqref{par_r_phi} (with $\Lambda_3= 1- \Lambda_1-\Lambda_2$)  we see that the volume element
transforms as
\begin{equation}
\ud V=
\ud\Lambda_1\,\ud\Lambda_2\ =
\frac{\sqrt{3}}{2}\,\times\,\chi_{\widehat{\mathbf{\Delta}}_2}\!\pt{r,\phi}\,\times\,r\,\ud r \,\ud \phi\quad.\nonumber
\end{equation}

\noindent The value of the radial variable $r$ is related to the purity of the \mbox{mixed state, $3/2\,
r^2 = \Tr\:\rho^2-1/3$, where $\Tr\:\rho^2=\Lambda_1^2+\Lambda_2^2+\Lambda_3^2$\ .}

\begin{figure}[ht]
\begin{center}
\includegraphics[width=\textwidth]{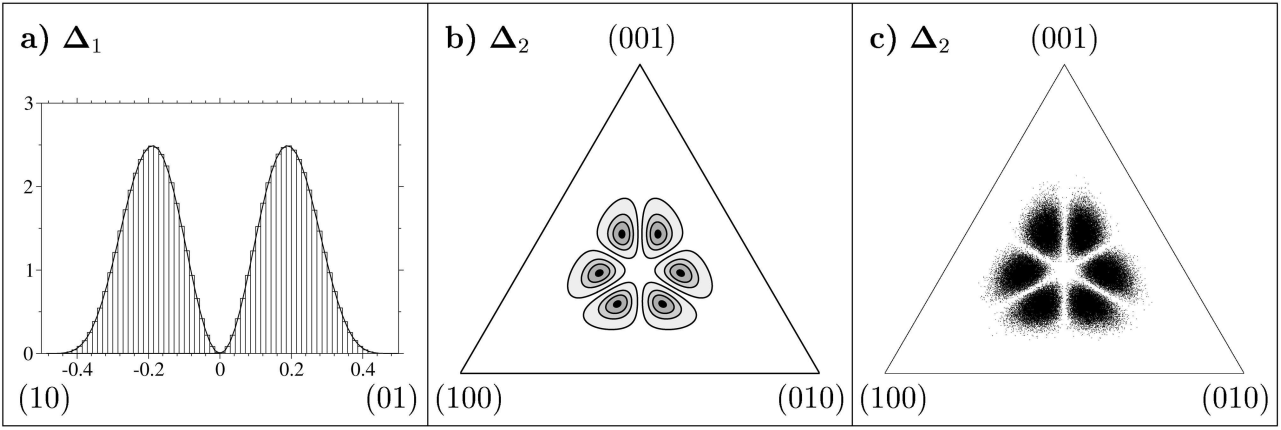}
\caption[]{Induced distributions in the simplices of eigenvalues obtained by partial trace of mixed states of $N\times N$ systems distributed according to \co{HS} measure.
{\bfseries (a)} $N=2$, $\mathbf{\widehat \Delta}_1=\pq{-1/2,1/2}$. Comparison of an histogram with theoretical prediction~\eqref{p28r} represented by solid line.
{\bfseries (b)} $N=3$. Contour lines of the distribution~\eqref{p327rp} in the simplex $\mathbf{\widehat \Delta}_2$.
{\bfseries (c)} $N=3$. Plot obtained numerically for $10^5$ random states of size $N^2=9$.
}%
\label{ch2}
\end{center}
\end{figure}

\noindent The constant $B^{(2)}_{N,N^3}$ follows from~\eqref{gen_mes_b}
            \begin{equation}
            B^{(2)}_{3,27} =
            \frac{80!}{12\cdot24!\cdot25!\cdot26!}\quad\cdot\nonumber
            \end{equation}
We are now in the position to compute for this case the explicit probability
distribution~\eqref{gen_meas}
\begin{equation}
P^{(2)}_{3,27}(\Lambda_1,\Lambda_2,\Lambda_3)\ = \ B^{(2)}_{3,27}\;
\chi_{\rule[0pt]{0pt}{7pt}\mathbf{\Delta}_2}\!\pt{\Lambda_1,\Lambda_2,\Lambda_3}\;\Lambda_1^{24}\;
\Lambda_2^{24}\;\Lambda_3^{24}\;{\pt{\Lambda_1-\Lambda_2}}^2\;{\pt{\Lambda_1-\Lambda_3}}^2
\;{\pt{\Lambda_2-\Lambda_3}}^2\quad,\label{p327l}%
\end{equation}
that in the polar plane reads
\begin{equation}
\widetilde{P}(r,\phi)\ = \
\frac{9}{64}\;\,\frac{80!}{24!\cdot25!\cdot26!}\;\,
\chi_{\widehat{\mathbf{\Delta}}_2}\!\pt{r,\phi}\;\,
r^6\sin^2\pt{3\phi}\;\,{\pt{\frac{1}{27}-\frac{1}{4}\:r^2+\frac{1}{4}\:r^3\cos\pt{3\phi}}}^{24}
\quad\cdot\label{p327rp}%
\end{equation}
The latter distribution is invariant under the transformations
$\phi\rightarrow-\phi$ and $\phi\rightarrow\phi+2k\pi/3,\
k\in\mathds{Z}$, as shown in FIG.~\ref{ch2}.
\section{Trace--non--increasing maps}%
\label{tnim} A generic state of a $N$--dimensional quantum system is
completely described once a positive, Hermitian and normalized
density matrix $\rho\in{\cal M}_N$ is given. In order to analyze and
classify the set of all possible physical operations on a quantum
system, we need to describe the set of {\it superoperators} $\Phi$
mapping ${\cal M}_N$ onto itself. The linearity of physical
operations can be expressed by forcing the superoperator $\Phi:{\cal
M}_N\mapsto{\cal M}_N$ to act as a matrix action on the ``vector''
$\rho$, that is
\begin{equation}
\rho^{\prime}=\Phi\rho
\qquad\text{or}\qquad\rho^{\prime}_{\supop{m\mu}{}}=
\Phi^{\phantom{\prime}}_{\supop{m\mu}{n\nu}}\rho^{\phantom{\prime}}_{\supop{n\nu}{}}\nonumber
\end{equation}
We use Einstein summation convention for indices appearing twice.
In order to map the domain ${\cal M}_N$ onto itself, the linear
super--operator $\Phi$ must fulfill these additional
\begin{quote}
\begin{PRS}{}\ \\[-4.5ex]
\label{prop_Phi}
\begin{Ventry}{}
\item[\ ] \ \\[-7.5ex]
\begin{subequations}
\begin{alignat}{4}
&\text{i})  &\qquad & \rho^{\prime}={\pt{\rho^{\prime}}}^{\dagger} &\qquad&\Leftrightarrow\qquad \Phi^{\phantom{\prime}}_{\supop{m\mu}{n\nu}}=\Phi^{\ast}_{\supop{\mu m}{\nu n}}&&\label{Phi_a}\\
&\text{ii}) &\qquad & \Tr\:\rho^{\prime}=\Tr\:\rho^{\phantom{\prime}}=1 &\qquad&\Leftrightarrow\qquad \Phi^{\phantom{\prime}}_{\supop{m m}{n\nu}}=\delta^{\phantom{\prime}}_{\supop{n\nu}{}}&&\label{Phi_b}\\
&\text{iii})&\qquad & \rho^{\prime}\geq0
&\qquad&\Leftrightarrow\qquad
\Phi^{\phantom{\prime}}_{\supop{m\mu}{n\nu}}\rho^{\phantom{\prime}}_{\supop{n\nu}{}}\geq
0 \qquad\text{when}\qquad\rho\geq 0\quad\cdot &&\label{Phi_c}
\end{alignat}
\end{subequations}
\end{Ventry}
\end{PRS}
\end{quote}
In particular, all the superoperators $\Phi$ fulfilling
property~\eqref{Phi_b} are called {\it trace preserving} (\co{TP}).
For any given $\Phi$, it proves convenient to introduce
the {\it dynamical matrix} $D_{\Phi}$, uniquely and
linearly obtained from the superoperator by reshuffling the indices~\cite{Sh50,SMR61}:
\begin{equation}
D^{\phantom{\prime}}_{\supop{m
n}{\mu\nu}}=\Phi^{\phantom{\prime}}_{\supop{m\mu}{n\nu}}\label{defD}\quad\cdot
\end{equation}
The dynamical matrix can be thought as a matrix on a bipartite $N
\times N$ quantum system $AB$, represented by means of an extended
Hilbert space ${\cal H}_{AB}$ of the same kind of the one in
Section~\ref{Mibptoms}, so that%
\  $D^{\phantom{\prime}}_{\supop{m n}{\mu\nu}}={}_A\!\left\langle
m\right|\otimes \!{}_B\left\langle
n\right|D_{\Phi}\left|\mu\right\rangle_A\otimes\left|\nu\right\rangle_B$.\\
In terms of $D_{\Phi}$, Properties~\ref{prop_Phi} can be
re--expressed as follows:
\begin{quote}
\begin{PRS}{}\ \\[-4.5ex]
\label{prop_D}
\begin{Ventry}{}
\item[\ ] \ \\[-7.5ex]
\begin{subequations}
\begin{alignat}{5}
&\text{i})  &\qquad & \rho^{\prime}={\pt{\rho^{\prime}}}^{\dagger} &\qquad&\Leftrightarrow\qquad D^{\phantom{\prime}}_{\supop{m n}{\mu \nu}}=D^{\ast}_{\supop{\mu \nu}{m n}}&\qquad&\text{so}&\qquad&D^{\phantom{\dagger}}_{\Phi}=D^{\dagger}_{\Phi}\label{D_a}\\
&\text{ii}) &\qquad & \Tr\:\rho^{\prime}=\Tr\:\rho^{\phantom{\prime}}=1 &\qquad&\Leftrightarrow\qquad D^{\phantom{\prime}}_{\supop{m n}{m \nu}}=\delta^{\phantom{\prime}}_{\supop{n\nu}{}}&&\text{so}&&\Tr_{A}D_{\Phi}=\mathds{I}_N\label{D_b}\\
&\text{iii})&\qquad & \rho^{\prime}\geq0
&\qquad&\Leftrightarrow\qquad
D^{\phantom{\prime}}_{\supop{mn}{\mu\nu}}\rho^{\phantom{\prime}}_{\supop{n\nu}{}}\geq
0 &&\text{when}&&\rho\geq 0 \label{D_c}
\end{alignat}
\end{subequations}
\end{Ventry}
\end{PRS}
\end{quote}
In the following we will focus on the set ${\cal CP}_N$ of maps
$\Phi:{\cal B}(\mathds{C}^{N})\mapsto{\cal B}(\mathds{C}^{N})$ that
are {\it completely positive}~\cite{St55}, that is on maps $\Phi$
such that for every identity map
$\mathds{I}_K:\mathds{C}^{K}\mapsto\mathds{C}^{K}$ the extended maps
$\Phi\otimes\mathds{I}_K:{\cal
B}(\mathds{C}^{N})\otimes\mathds{C}^K\mapsto{\cal
B}(\mathds{C}^{N})\otimes\mathds{C}^K$ are positive (here ${\cal
B}(\mathds{C}^{N})$ is the Banach space of bounded linear operators
on $\mathds{C}^{N}$). A very important property characterizes the
dynamical matrix $D_{\Phi}$ of a completely positive (\co{CP}) map $\Phi$:
\begin{quote}
\begin{TT}[Choi~\cite{Ch75}]{:} \label{Choi}
A linear map $\Phi$ is completely positive if and only if the
corresponding dynamical matrix $D_{\Phi}$ is positive.
\end{TT}
\end{quote}
Note that property~\eqref{D_c} holds true in general once that the
positivity of the matrix $D_{\Phi}$ is proved for product states of
$\mathds{C}^{N}\otimes\mathds{C}^{N}$, as stated in the
Jamio{\l}kowski Theorem~\cite{Ja72}. This property is implied by
complete positivity, that is a stronger condition. If we combine
equations~(\ref{D_a}--\ref{D_b}) with the complete positivity, we
observe that for any given $\Phi\in{\cal CP}_N^{\co{TP}}$ (i.e. the
set of completely positive   trace preserving maps), its {\it
rescaled dynamical matrix} $\rho_{\Phi}\coleq D_{\Phi}/N$ possesses
the three properties of Hermiticity, positivity and normalization.
Thus $\rho_{\Phi}$ belongs to the set ${\cal
M}_{N^2}^{\phantom{\mathds{I}}}$ of density matrices of size $N^2$.
This leads to the celebrated Jamio{\l}kowski
isomorphism~\cite{Ja72}, that maps the entire set ${\cal
CP}_N^{\co{TP}}$ of quantum maps onto a proper
$(N^4-N^2)$--dimensional subset of the set ${\cal
M}_{N^2}^{\phantom{\mathds{I}}}$ of quantum states on extended
system. This subset, denoted by ${\cal M}_{N^2}^{\mathds{I}}$\ ,
contains all states $\rho$ such that $\Tr_A\rho=\mathds{I}_N$. The
$N^2$ constraints reducing the dimension of ${\cal
M}_{N^2}^{\phantom{\mathds{I}}}$ come from equation~\eqref{D_b}.
With the aim of removing these $N^2$ constraints, we introduce a
family of linear maps:
\begin{quote}
\begin{DDD}{}\ \\[-5.5ex]
\begin{Ventry}{}\label{tnimaps2}
    \item[] A linear positive map $\Phi$ is called
\textit{trace--non--increasing} (\co{TNI}), if
 ${\rm Tr} \;\Phi(\rho) \leq {\rm Tr}\; \rho = 1$ for any $\rho\in{\cal M}_N$\ .
\end{Ventry}
\end{DDD}
\end{quote}
We state now a Lemma that makes a link between \co{CP--TNI} maps and
their images given by Jamio{\l}kowski isomorphism, that is the set
of their (rescaled) dynamical matrices:
\begin{quote}
\begin{LLL}{}\ \label{Lemma2}\\
For any given $\Phi\in{\cal CP}^{\text{\co{TNI}}}_N$, its
rescaled dynamical matrix $\sigma_{\Phi}\coleq D_{\Phi}/N$ is  Hermitian
(according to~\eqref{D_a}), positive (as in the statement of Choi
Theorem) and fulfills the following constraint:
\begin{equation}
{\text{ii}}^{\prime}) \qquad
\Tr\:\Phi(\rho)=\Tr\:\rho^{\prime}\leq\Tr\:\rho^{\phantom{\prime}}=1
\qquad\Leftrightarrow\qquad
\Tr_A\;\sigma_{\Phi}\leq\frac{\mathds{I}_N}{N}\label{D_l2}\quad\cdot
\end{equation}
\end{LLL}
\end{quote}
\noindent For a proof of this Lemma, see Appendix~\ref{app_A2}.
Note that the above constraint is a kind of a relaxation
of~\eqref{D_b}. The {\it reduced and rescaled dynamical matrix} $\Tr_A \sigma_{\Phi}$ belongs then to a subset of the set ${\cal M}_N^{\text{\co{sub}}}$ defined in~\eqref{convhull}. To specify this subset consider the following set of {\it sub--tracial
states}
\begin{equation}
{\cal M}_N^{\Box} \coleq\bigg\{\sigma\in{\cal
M}_N^{\text{\co{sub}}}\ :\
\sigma\leq\frac{\mathds{I}_N}{N}\bigg\}=\bigg\{\sigma\in{\cal
M}_N^{\text{\co{sub}}}\ :\
\max\pq{\text{\co{EV}}\pt{\sigma}}\leq\frac{1}{N} \bigg\}
\quad, \label{M_square}%
\end{equation}
where $\text{\co{EV}}\pt{\sigma}$ denotes the set of eigenvalues of
$\sigma$. Note that the set ${\cal M}_N^{\Box}$ coincides, up to
rescaling by $N$, with the set of positive Hermitian operators $E_i$
fulfilling $E_i\leq\mathds{I}_N$\ . A collection of such operators,
which satisfies \mbox{the relation $\sum_i E_i = \mathds{I}_N$\ ,}
is called \co{POVM} (\co{P}ositive \co{O}perator \co{V}alued
\co{M}easure) and plays a crucial role in the theory of quantum
measurement~\cite{BZ06}.

Definition~\eqref{M_square} allows us to rewrite
condition~\eqref{D_l2} as
\begin{equation}
\Phi\in{\cal CP}^{\text{\co{TNI}}}_N \qquad\Leftrightarrow\qquad
\Tr_A\;\sigma_{\Phi}\in{\cal M}_N^{\Box} \label{D_l2_fin}\quad\cdot
\end{equation}
\noindent Finally, we can summarize the result of
equation~\eqref{D_l2_fin}, in the next
\begin{quote}
\begin{PPP}{}\ \\[-5.5ex]
\begin{Ventry}{}\label{repr1}
    \item[] Every trace non increasing map $\Phi\in{\cal
CP}_N^{\co{TNI}}$ can be represented by a sub--tracial state
$\check{\sigma}_{\Phi}\in{\cal M}_N^{\Box}\subset{\cal
M}_N^{\text{\co{sub}}}$, whose explicit expression is given in terms
of the rescaled dynamical matrix $\sigma_\Phi$ of Lemma~\ref{Lemma2}
by
\begin{equation}
\check{\sigma}_{\Phi}\coleq\Tr_A\;\sigma_{\Phi}=\frac{1}{N}\Tr_A\;D_{\Phi}\quad\nonumber\cdot
\end{equation}
Moreover, the rescaled dynamical matrix $\sigma_\Phi$ offers itself
another representation of $\Phi$ into an $N^4$--dimensional proper
subset of ${\cal M}_{N^2}^{\text{\co{sub}}}$\ , as one can derive
from the right hand side of~\eqref{D_l2} that
\begin{equation}
\Tr\:\sigma_{\Phi}=\Tr_{AB}\;\sigma_{\Phi}=\Tr_{B}\;\pt{\Tr_{A}\;\sigma_{\Phi}}
\leq\Tr_{B}\;\pt{\mathds{I}_N/N}=
1\quad\nonumber\cdot
\end{equation}
In the next Definition~\ref{repr2}, we will denote this set with ${\cal M}_{N^2}^{\boxplus}$.
\end{Ventry}
\end{PPP}
\begin{DDD}{}\ \\[-5.5ex]
\begin{Ventry}{}\label{repr2}
    \item[]\ \\[-5.5ex]
\begin{equation}
{\cal M}_{N^2}^{\boxplus}\coleq\bigg\{\sigma\in{\cal M}_{N^2}^{\text{\co{sub}}}\ :\
\Tr_{A}\;\sigma\in{\cal M}_{N^{\phantom{2}}}^{\Box} \bigg\}\quad\nonumber.
\end{equation}
\end{Ventry}
\end{DDD}
\end{quote}
\noindent To clarify the notation used we collect the sets of states and maps considered in
table~\ref{tab:table}.
\begin{table}[htbp]
\begin{center}
\begin{tabular}{||p{45mm}||*{3}{c|}|}
\hline
\rule[-2.3ex]{0pt}{6.6ex}& Map $\Phi$ & $\newatop{\displaystyle \text{Rescaled dynamical}}{\displaystyle \rule[0ex]{0pt}{1.7ex}\text{matrix }\sigma_{\Phi}}$ & $\newatop{\displaystyle \text{Reduced and rescaled}}{\displaystyle \rule[0ex]{0pt}{2.1ex}\text{dynamical matrix }\check{\sigma}_{\Phi}}$\\
\hline \hline
\rule[-2.5ex]{0pt}{6.5ex}\co{CP} \ \ \co{T}race \co{P}reserving maps & ${\cal CP}_N^{\co{TP}}$ & ${\cal M}_{N^2}
^{\mathds{I}}=\pt{{\cal M}_{N^2}^{\phantom{\text{\co{sub}}}}\!\!\cap{\cal M}_{N^2}^{\boxplus}}$ &
$\frac{\mathds{I}_N}{N}\in\pt{{\cal M}_{N^{\phantom{2}}}^{\phantom{\text{\co{sub}}}}\!\!\!\cap{\cal M}_{N^{\phantom{2}}}^{\Box}}$ \\
\hline
\rule[-2.5ex]{0pt}{6.5ex}\co{CP} \ \ \co{T}race \co{N}on \co{I}ncreasing maps &
${\cal CP}_N^{\co{TNI}}$ & ${\cal M}_{N^2}^{\boxplus}\subset{\cal M}_{N^2}^{\text{\co{sub}}}$ & ${\cal M}_{N^{\phantom{2}}}^{\Box}\subset{\cal M}_{N^{\phantom{2}}}^{\text{\co{sub}}}$ \\
\hline
\end{tabular}
\end{center}
\caption{The two kinds of \co{CP}-\co{TP} and \co{CP}--\co{TNI} maps
analyzed in this Section are mapped in the set of their
correspondent superoperators $\Phi$, the set of their rescaled
dynamical matrices $\sigma_{\Phi}$, and the set of their representative
sub--tracial states $\check{\sigma}_{\Phi}$. Inclusion relations
between the sets under consideration are explicitly shown in the table.}
\label{tab:table}
\end{table}

\begin{figure}[ht]
\begin{center}
\includegraphics[width=\textwidth]{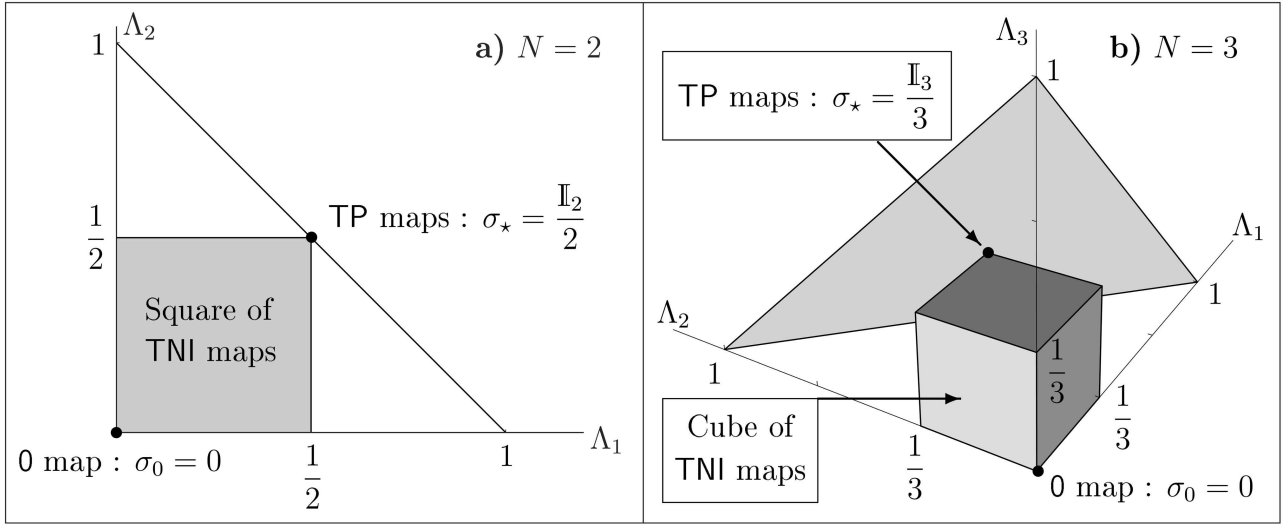}
\caption[]{The set of \co{CP} trace-non-increasing maps
$\Phi\in{\cal CP}^{\co{TNI}}_N$ represented, by means of the
Jamio{\l}kowski isomorphism, as the subset ${\cal M}_{N^{\phantom{2}}}^{\Box}$ of the set of subnormalized states
${\cal M}_N^{\rm sub}$, in terms of the reduced and rescaled dynamical
matrices $\check{\sigma}_{\Phi}=\Tr_A\sigma_{\Phi}$. The  eigenvalues of these matrices $\check{\sigma}_{\Phi}$ span an $N$--dimensional cube inscribed
into the set of subnormalized states of size $N$, here
 plotted for  {\bfseries (a)} $N=2$ and {\bfseries (b)} $N=3$.}%
\label{cube}
\end{center}
\end{figure}

In the space of the $N$--eigenvalues of  $N\times N$ positive
Hermitian matrices, the set ${\cal M}_N^{\Box} $ of sub--tracial
states defined in~\eqref{M_square} consists of a cube inscribed into
the set ${\cal M}_N^{\text{\co{sub}}}$ of subnormalized states of
size $N$. Such a multi--dimensional cube has a vertex in the origin,
is oriented along axes, and touches the simplex of quantum states
${\cal M}_N$ in a single point $\mathds{I}_N/N$, as shown in FIG.~\ref{cube} for $N=2$ and $N=3$. Observe that
every $\check{\sigma}_{\Phi}$ of ${\cal
M}_{N^{\phantom{2}}}^{\phantom{\Box}}$ can be rescaled in order
to let it belong to the cube of ${\cal
M}_{N^{\phantom{2}}}^{\Box}$. In terms of maps it means that
every $\Phi\in{\cal CP}_N^{\phantom{\co{TP}}}$ can be
mixed with the $0$ map in order to become
trace--non--increasing.

The set ${\cal M}_{N^2}^{\boxplus}$ of \co{CP--TNI} maps has $N^4$ dimensions. It
contains the $N^4-N^2$ dimensional set ${\cal M}_{N^2}^{\mathds{I}}$ of \co{CP--TP} maps, which
satisfy ${\rm Tr} \;\Phi(\rho) = {\rm Tr}\; (\rho)$, as a subset.
The set ${\cal
CP}^{\text{\co{TP}}}_N$ forms extremal points in  ${\cal
CP}^{\text{\co{TNI}}}_N$.

\section{Volume of the set of \co{CP TNI} maps}%
\label{Vol_TNIM}
In Section~\ref{tnim}, the Jamio{\l}kowski isomorphisms allowed us
to establish links between superoperators $\Phi\in{\cal
CP}_N^{\co{TNI}}$, rescaled dynamical matrices ${\sigma}_{\Phi}=\Tr_A\;\sigma_{\Phi}\in{\cal M}_{N^2}^{\boxplus}$\ , and sub--tracial states $\check{\sigma}_{\Phi}\in{\cal
M}_N^{\Box}$. In this Section we will make use of this representation aiming to define a measure in ${\cal
CP}_N^{\co{TNI}}$ and compute its volume.

We start defining the \co{H}ilbert \co{S}chmidt measure $\ud \mu_{\text{\co{HS}}}^{\boxplus}$ on the space of trace non--increasing maps $\Phi\in{\cal CP}^{\co{TNI}}_N$ as the \co{HS} measure of their representative reduced dynamical matrices ${\sigma}_{\Phi}\in{\cal M}_{N^2}^{\boxplus}\subset{\cal M}_{N^2}^{\text{\co{sub}}}$. Such a measure is analogous to the one introduced in~(\ref{vol_el_sub}--\ref{dimuch}), restricted from ${\cal M}_{N^2}^{\text{\co{sub}}}$ to ${\cal M}_{N^2}^{\boxplus}$\ , and reads
\begin{equation}
\ud
\mu_{\text{\co{HS}}}^{\boxplus}\pt{\Phi}=
\ud \mu_{\text{\co{HS}}}^{\text{\co{sub}}}\pt{\sigma_\Phi}\bigg|_{{\cal M}_{N^2}^{\boxplus}\subset{\cal M}_{N^2}^{\text{\co{sub}}}}=\frac{1}{N^2!}\;\ud\nu^{\text{\co{sub}}}
(\Lambda_{1^{\phantom{2}}},\ldots,\Lambda_{N^2})\bigg|_{{\cal M}_{N^2}^{\boxplus}\subset{\cal M}_{N^2}^{\text{\co{sub}}}}\times\ud\nu^{\text{\co{\,Haar}}}\quad
\cdot\label{vol_el_tni}
\end{equation}
Here $\Lambda_{1^{\phantom{2}}},\ldots,\Lambda_{N^2}$ denote the eigenvalues of the
matrix ${\sigma}_{\Phi}$ of size $N^2$.

The restriction of the space from ${\cal M}_{N^2}^{\text{\co{sub}}}$
to ${\cal M}_{N^2}^{\boxplus}$ does not affect the Haar measure on
the entire complex flag manifold $Fl^{(N^2)}_{\mathds{C}}$, given
by the coset space $\mathrm{U}(N^2)/{\pq{\mathrm{U}(1)}}^{N^2}$ of
equivalence classes of unitary matrices of eigenstates of $\sigma_{\Phi}$.
The volume of this flag manifold follows from~\eqref{flagman}
\begin{equation}
\mathrm{Vol}\pq{Fl^{(N^2)}_{\mathds{C}}}=\int_{Fl^{(N^2)}_{\mathds{C}}}\ud\nu^{\text{\co{\,Haar}}}=
\frac{{(2\pi)}^{N^2(N^2-1)/2}}{1\,!\;2\,!\;\cdots(N^2-1)\,!}\label{flagmantni}
\quad\cdot
\end{equation}
In order to parallel~\eqref{volqs} and~\eqref{volsub} we need to compute the volume spanned by the eigenvalues of matrices contained in
${\cal M}_{N^2}^{\boxplus}$ according to the measure $\ud\nu^{\text{\co{sub}}}
(\Lambda_{1^{\phantom{2}}},\ldots,\Lambda_{N^2})$, multiply it by the volume~\eqref{flagmantni} of the flag manifold $Fl^{(N^2)}_{\mathds{C}}$ and divide the result by $N^2!$ as in~\eqref{vol_el_tni}. Performing this task one obtains
\begin{equation}
\mathrm{Vol}_{\textsf{\:HS}}\pt{{\cal CP}^{\co{TNI}}_N}=\frac{1}{N^2!}\:\mathrm{Vol}\pq{\text{\co{EV}}\pt{{\cal M}_{N^2}^{\boxplus}}}\times\mathrm{Vol}\pq{Fl^{(N^2)}_{\mathds{C}}}=\frac{1}{N^2!}\:
\int_{\text{\co{EV}}\pt{{\cal M}_{N^2}^{\boxplus}}}\ud\nu^{\text{\co{sub}}}(\Lambda_{1^{\phantom{2}}},\ldots,\Lambda_{N^2})
\times\mathrm{Vol}\pq{Fl^{(N^2)}_{\mathds{C}}}
\quad\cdot\label{voltni}
\end{equation}
Note that the operation of partial trace maps the set ${\cal M}_{N^2}^{\boxplus}$ into ${\cal M}_{N^{\phantom{2}}}^{\Box}$. Assuming that mixed states from ${\cal M}_{N^2}^{\boxplus}\subset{\cal M}_{N^2}^{\text{\co{sub}}}$ are distributed according to the \co{HS} measure, we need to analyze the measure induced in ${\cal M}_{N^{\phantom{2}}}^{\Box}\subset{\cal M}_{N^{\phantom{2}}}^{\text{\co{sub}}}$ by partial trace.

In Section~\ref{ptorhsdms} we described how the normalized \co{HS}--distribution on ${\cal M}_{NK}$, whose expression is equivalent to the induced distribution $P_{NK,NK}^{(2)}\pt{\Lambda_1,\ldots,\Lambda_{NK}}$ of equation~\eqref{gen_meas}, is mapped by partial tracing into the distribution $P_{N,NK^2}^{(2)}\pt{\Lambda_1,\ldots,\Lambda_{N}}$. In particular, for $K=N$, the \co{HS}--distribution on ${\cal M}_{N^2}$ is mapped into $P_{N,N^3}^{(2)}\pt{\Lambda_1,\ldots,\Lambda_{N}}$.

In a completely equivalent way, the measure
$\ud\nu^{\text{\co{sub}}}(\Lambda_1,\ldots,\Lambda_{N^2})=\ud\nu_{N^2}(\Lambda_1,\ldots,\Lambda_{N^2})$
of equation~\eqref{vol_el_tni} is transformed by partial tracing
into the measure $\ud\nu_{N^3}(\Lambda_1,\ldots,\Lambda_N)$,
and~\eqref{voltni} yields
\begin{align}
\mathrm{Vol}_{\textsf{\:HS}}\pt{{\cal CP}^{\co{TNI}}_N}& =\frac{1}{N^2!}\: \int_{\text{\co{EV}}\pt{{\cal
M}_{N^2}^{\boxplus}}}\ud\nu_{N^2}(\Lambda_{1^{\phantom{2}}},\ldots,\Lambda_{N^2})
\times\mathrm{Vol}\pq{Fl^{(N^2)}_{\mathds{C}}}
 =\frac{\mathrm{Vol}\pq{Fl^{(N^2)}_{\mathds{C}}}}{N^2!}\:
\int_{\text{\co{EV}}\pt{{\cal
M}_{N}^{\Box}}}\ud\nu_{N^3}(\Lambda_{1^{\phantom{2}}},\ldots,\Lambda_{N})
\quad\label{enne2}.%
 \intertext{Making use of~\eqref{nu_K} and~\eqref{flagmantni} we arrive at}
\mathrm{Vol}_{\textsf{\:HS}}\pt{{\cal CP}^{\co{TNI}}_N} & =\frac{{(2\pi)}^{N^2(N^2-1)/2}}{1\,!\;2\,!\;\cdots N^2\,!}\:
\int \prod_{i=1}^N \,\Theta\pt{\frac{1}{N}- \Lambda_i}
\quad\times\quad \prod_{i=1}^N
\,\Theta(\Lambda_i)\:\Lambda_i^{N^3-N} \:\prod_{i<j}
\pt{\Lambda_i-\Lambda_j}^2\;\ud\Lambda_1\,\ldots\,\ud\Lambda_N
\quad,\label{enne3}
\end{align}
where we have made explicit the domain of integration using the
Heaviside step function $\Theta$. Last integral can be computed
using next Lemma~\ref{Lemma3}, whose proof is in
Appendix~\ref{app_A3}.
\begin{quote}
\begin{LLL}{}\ \label{Lemma3}\\
On the $N$--dimensional cube $\Box_N\coleq
{\pq{0,\frac{1}{N}}}^{N}$, consider the one parameter family of
measures $\ud\nu_K^{\Box}(\Lambda_1,\ldots,\Lambda_N)$
\begin{equation}
\ud\nu_K^{\Box}(\Lambda_1,\ldots,\Lambda_N)=\,\prod_{i=1}^N\Theta\pt{\frac{1}{N}-
\Lambda_i}\:\Theta(\Lambda_i)\:\Lambda_i^{K-N}\prod_{i<j}
\pt{\Lambda_i-\Lambda_j}^2\;\ud\Lambda_1\,\ldots\,\ud\Lambda_N\quad\cdot
\label{nu_Kcubo}
\end{equation}
labeled by any integer $K\geq N$. Then the volume
$\nu_K^{\Box}\pt{\Box_N}$ reads
\begin{equation}
\int_{\Box_N}\ud\nu_K^{\Box}(\Lambda_1,\ldots,\Lambda_N) =
\frac{I(K-N+1,1,1,N)}{N^{KN}}\quad , \label{HJ5}
\end{equation}
 where $I(K-N+1,1,1,N)$ is the Selberg's integral~\cite{Me91}
\begin{equation}
 I(K-N+1,1,1,N)=\prod_{j=1}^{N}\frac{\Gamma\pt{1+j}\:\Gamma\pt{K-N +j}\:\Gamma\pt{
j}}{\Gamma\pt{2}\:\Gamma\pt{K +j}}\quad\cdot\label{Selb}
\end{equation}
\end{LLL}
\end{quote}
\noindent Finally, using Lemma~\ref{Lemma2}, equation~\eqref{enne3}
yields the final formula for the \co{HS} volume of the set of \co{TNI} maps
\begin{align}
\mathrm{Vol}_{\textsf{\:HS}}\pt{{\cal CP}^{\co{TNI}}_N}&
=\frac{{(2\pi)}^{N^2(N^2-1)/2}}{1\,!\;2\,!\;\cdots N^2\,!}\:
\int_{\Box_N}\ud\nu_{N^3}^{\Box}(\Lambda_1,\ldots,\Lambda_N)\nonumber\\%
& =\frac{{(2\pi)}^{N^2(N^2-1)/2}}{1\,!\;2\,!\;\cdots N^2\,!}\:
\prod_{j=1}^{N}\frac{\Gamma\pt{1+j}\:\Gamma\pt{N^3-N +j}\:\Gamma\pt{
j}}{N^{N^3}\:\Gamma\pt{N^3 +j}}
\end{align}
    Equation~\eqref{enne2} implies that  the \co{HS} volume of
    the set of \co{CP--TNI} maps is proportional to the volume of the
    set of sub--tracial states ${\cal M}_N^{\Box}$ of~\eqref{M_square} computed
    according to the induced measure
    $\ud\nu_{N^3}(\Lambda_{1^{\phantom{2}}},\ldots,\Lambda_{N})$. In
    particular Lemma~\ref{Lemma3} allows us to compute such a
    volume, which is in turn proportional to the volume spanned by all possible elements
    of a \co{POVM}, according to any given induced measure
    $\ud\nu_{K}(\Lambda_{1^{\phantom{2}}},\ldots,\Lambda_{N})$.
    Moreover, as we explained in the Remark of page~\pageref{rem1}, also
    the volume of all sub--normalized states ${\cal
    M}_N^{\text{\co{sub}}}$ can be computed according to the any induced measure.
    For comparison, we present below their ratio for any value of parameter $K$,
    reminding the reader that in the former calculations we set $K=N^3$.
\begin{equation}
\frac{\mathrm{Vol}_{N,K}\pt{{\cal
M}_N^{\Box}}}{\mathrm{Vol}_{N,K}\pt{{\cal M}_N^{\text{\co{sub}}}}}
=\frac{NK}{N^{NK}} \;C_N^{(K-N+1,2)}\times I(K-N+1,1,1,N)
=\frac{\pt{NK}\text{{\large
!}}}{N^{NK}}\prod_{j=1}^{N}\frac{\Gamma\pt{j}}{\Gamma\pt{K+j}}\quad\cdot\nonumber
\end{equation}

\section{Extremal \co{CP TNI} maps}%
\label{ECPTNIm}
We start this Section by introducing another representation for
\co{CP}--\co{TNI} maps $\Phi:\rho\longmapsto\rho^{\prime}$,
alternative to the one described in Section~\ref{tnim}, and
summarized in the next Lemma~\ref{Lemma4} (for a proof see
Appendix~\ref{app_A4}).
\begin{quote}
\begin{LLL}{}\ \label{Lemma4}\\
For any given \co{CP}--\co{TNI} map $\Phi$, its action on density
matrices $\rho\in{\cal M}_N$ can be described by using a discrete
family of $N\times N$ operators $A_\mu$, whose number $S$ does not
exceed $N^2$. The explicit action of the map $\Phi$ reads
\begin{subequations}
\label{ks_tnim}
\begin{equation}
\Phi\pt{\rho} =\sum_{\mu=1}^S
A_\mu^{\phantom{\dagger}}\:\rho\:A_\mu^{\dagger}\quad\label{ks_tnim1}
\end{equation}
and the matrices ${A_\mu}$ fulfill
\begin{equation}
\sum_{\mu=1}^S A_\mu^{\dagger}\:A_\mu^{\phantom{\dagger}}
=N\:\check{\sigma}_{\Phi}^{\text{\co{T}}}\quad,\label{ks_tnim2}%
\end{equation}
\end{subequations}
where the sub--tracial state $\check{\sigma}_{\Phi}\in{\cal
M}_N^{\Box}$ is given by the reduced and rescaled dynamical matrix
correspondent to the map $\Phi$, as in the Proposition of
pag.~\pageref{repr1}, and $\sigma^{\text{\co{T}}}$ means
\emph{transpose of} $\sigma$. Conversely, if $S\leq N^2$ operators
$A_\mu$ of size $N\times N$ fulfill
\begin{subequations}
\label{ks_tnim_inv}
\begin{equation}
 \frac{1}{N}\sum_{\mu=1}^S A_\mu^{\dagger}\:A_\mu^{\phantom{\dagger}}
=\check{\sigma}\in{\cal
M}_N^{\Box}\quad,\label{ks_tnim_inv1}%
\end{equation}
then
\begin{equation}
\Phi\pt{\rho} \coleq\sum_{\mu=1}^S
A_\mu^{\phantom{\dagger}}\:\rho\:A_\mu^{\dagger}\quad\label{ks_tnim_inv2}
\end{equation}
defines a \co{CP}--\co{TNI} map.
\end{subequations}
\end{LLL}
\end{quote}
When $\Phi$ is a \co{CP}--\co{TP} map, that is when
$\check{\sigma}_{\Phi}^{\phantom{\text{\co{T}}}}=\check{\sigma}_{\Phi}^{\text{\co{T}}}=\mathds{I}_N/N$,
then equations~\eqref{ks_tnim} define the usual
\co{K}raus--\co{S}tinespring representation and the operators
$A_\mu$ are said \emph{Kraus operators}. In this sense
Lemma~\ref{Lemma4} can be seen as an extension of the
\co{KS}--representation.

\begin{figure}[ht]
\begin{center}
\includegraphics[width=87.742mm]{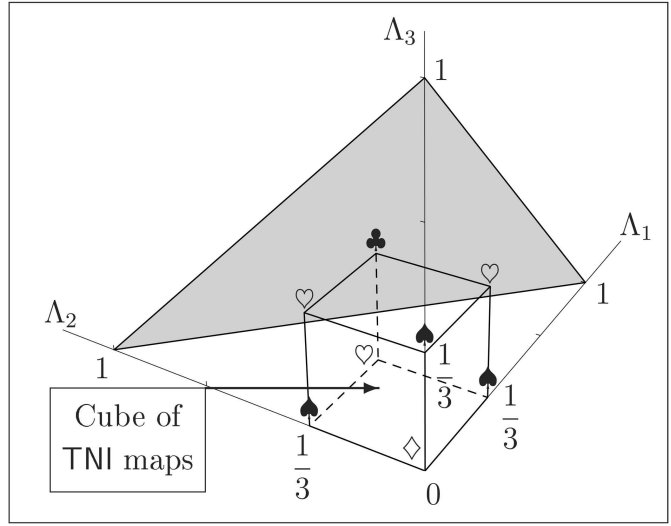}
\caption[]{Set of \co{CP}--\co{TNI} maps with extremal points
marked. In particular $\pt{\clubsuit}$ marks the \co{CP}--\co{TP}
map $\Phi_{3}=\mathds{I}_3/3$, $\pt{\diamondsuit}$ marks the $0$ map
$\Phi_{0}=0$, $\pt{\heartsuit}$ and $\pt{\spadesuit}$ marks maps
$\Phi_{2}$, respectively
$\Phi_{1}$.}%
\label{cube_suites}
\end{center}
\end{figure}

We aim to use the previous Lemma on a particular sub--class of
\co{CP}--\co{TNI} maps, represented by very trivial sub--tracial
states, as in the following
\begin{quote}
\begin{DDD}{}\ \\[-5.5ex]
\begin{Ventry}{}\label{k_tnim_def}
    \item[] We say that a map $\Phi\in{\cal CP}^{\co{TNI}}_N$ is
    $k$\emph{--extremal} if its representative sub--tracial state
    $\check{\sigma}_{\Phi}\in{\cal M}_N^{\Box}$ is diagonal and
    possesses exactly $k$ non vanishing entries, all equal to $1/N$.
    Those sub--tracial states represent maximally mixed states in
    $k$--dimensional subspace.
    In the following such maps will be denoted by $\Phi_k$.
\end{Ventry}
\end{DDD}
\end{quote}
 Let $\Omega_N$ denote the ordered set of positive integer less or
    equal to $N$, and let $\mathcal{P}\pt{\Omega_N}$ be the set ``\emph{sorted parts of }
    $\Omega_N$'', that is the set of all possible ordered collections of elements of $\Omega_N$.
    Consider the partition
\begin{equation}
\mathcal{P}\pt{\Omega_N}=\;
 \bigcup_{k=0}^N\mathcal{P}_k\pt{\Omega_N}\quad,\label{part_omega}
\end{equation}
where each $\mathcal{P}_k\pt{\Omega_N}$ contains exactly $k$
elements, $\mathcal{P}_0\pt{\Omega_N}=\pg{\emptyset}$ and
$\mathcal{P}_N\pt{\Omega_N}=\pg{\Omega_N}$. Thus it proves
convenient to use the natural isomorphism
\begin{equation}
\zeta\in\mathcal{P}_k\pt{\Omega_N}\quad \text{\Large{$\sim$}}\quad
\check{\sigma}_{\zeta}=\frac{1}{N}\sum_{\ell\in\zeta}\ket{\ell}\bra{\ell}
\label{k_isomorph}
\end{equation}
for labeling the set of $k$\emph{--extremal} \co{CP}--\co{TNI} maps:
see Figure~\ref{cube_suites} for an example with $N=3$, on which
\begin{equation}
 \mathcal{P}_3\pt{\Omega_3}=\pg{\Omega_3}=\pg{\pg{1,2,3}}\quad,\quad
 \mathcal{P}_2\pt{\Omega_3}=\pg{\pg{1,2},\pg{1,3},\pg{2,3}}\quad,\quad
 \mathcal{P}_1\pt{\Omega_3}=\pg{\pg{1},\pg{2},\pg{3}}\quad\text{and}\quad
 \mathcal{P}_0\pt{\Omega_3}=\pg{\emptyset}\quad.\nonumber
\end{equation}
The vectors $\ket{\ell}$ are meant to be orthonormal .
 Figure~\ref{cube_suites} illustrates another geometrical meaning
of the integer number $k$ labeling a $k$--extremal map: it is
proportional to the \emph{taxi distance} in $\mathds{R}^N$ between
$\vec{0}$ and the vector of diagonal entries of
$\check{\sigma}_{\zeta}$, that is the minimal number of sides
connecting $0$ to $\check{\sigma}_{\zeta}$ in the plot of
eigenvalues.

From~\eqref{k_isomorph} we earn that $\check{\sigma}_{\zeta}$ is
essentially a projector operator rescaled  by $N$,
$\check{\sigma}_{\zeta}=P_{\zeta}/N$: we denote as ${\cal
M}_{\zeta}$ the image of ${\cal M}_{N}$ obtained by projection,
\begin{equation}
{\cal M}_{\zeta}\coleq
\bigcup_{\rho\in{\cal M}_N}P_{\zeta}\:\rho
\:P_{\zeta}\qquad\subset{\cal M}_N^{\text{\co{sub}}}
 \label{M_zeta}\quad.
\end{equation}
Using the isomorphism~\eqref{k_isomorph}, Lemma~\ref{Lemma4} can be
re--phrased as follows:
\begin{quote}
\begin{LLL}{}\ \label{Lemma5}\\
For any given $k$-extremal \co{CP}--\co{TNI} map $\Phi_k$, that is
for any given $\zeta\in\mathcal{P}_k\pt{\Omega_N}$, its action on
density matrices $\rho\in{\cal M}_N$ can be described by using a
discrete family of $N\times N$ operators $A_\mu$, whose number $S$
does not exceed $N^2$. The explicit action of the map $\Phi_k$ reads
\begin{subequations}
\label{k_tnim}
\begin{equation}
\Phi\pt{\rho} =\sum_{\mu=1}^S
A_\mu^{\phantom{\dagger}}\:\rho\:A_\mu^{\dagger}\quad\label{k_tnim1}
\end{equation}
and the matrices ${A_\mu}$ fulfill
\begin{equation}
\sum_{\mu=1}^S A_\mu^{\dagger}\:A_\mu^{\phantom{\dagger}}
=N\:\check{\sigma}_{\zeta}\quad.\label{k_tnim2}%
\end{equation}
\end{subequations}
\end{LLL}
\end{quote}
Simply using the previous Lemma~\ref{Lemma5} we deduce the following
\begin{quote}
\begin{PRS}{}\ \\[-1.5ex]
\label{prop_k} Let $\Phi_k$ be a $k$-extremal \co{CP}--\co{TNI} map,
represented by some $\check{\sigma}_{\zeta}$, with
$\zeta\in\mathcal{P}_k\pt{\Omega_N}$. Then it holds true:
\begin{Ventry}{\textbf{$3$)}}
\item[$\mathbf{1)}$] when $k=N$, $\Phi_N$ belongs to ${\cal CP}^{\co{TP}}_N$. When $k<N$ the trace
preserving condition becomes $\rho$--dependent: nevertheless
$\Phi_k\in{\cal CP}^{\co{TNI}}_N$ acts as a \co{T}race
\co{P}reserving map on $\rho\in{\cal M}_\zeta$\ ;
\item[$\mathbf{2)}$] acting on the maximally mixed state $\rho_\star\coleq\mathds{I}_N/N$
reveals the parameter $k$, being
$\Tr\:\pq{\Phi_k\pt{\rho_\star}}=k/N$;
\item[$\mathbf{3)}$] ${\cal
M}_N^{\text{\co{sub}}}\supseteq\Phi_k\pt{\rho_\star}=\pt{k/N}\omega$,
for some $\omega\in{\cal M}_N$. When $k=N$ and $\Phi$ is
\emph{unital}, then $\omega=\rho_\star$\ .
\end{Ventry}
\end{PRS}
\end{quote}
\newpage
\textbf{Proof of Properties~\ref{prop_k}:}\\[-1ex]

\noindent$\mathbf{1)}$\quad  Tracing both sides of
equation~\eqref{k_tnim1}, using the cyclicity property of the trace,
and inserting~\eqref{k_tnim2}, one obtains $\displaystyle
\Tr\:\pq{\Phi_k\pt{\rho}}=\Tr\:\pt{N\,\check{\sigma}_{\zeta}\,\rho}=\Tr\:\pt{P_{\zeta}\,\rho}$,
where $P_\zeta$ is the projector of equation~\eqref{M_zeta}. Then
using the property of projector and ciclicity, together with
definition~\eqref{M_zeta}, one gets
\begin{equation}
\Tr\:\pq{\Phi_k\pt{\rho}}=\Tr\:\pt{P^2_{\zeta}\,\rho}=\Tr\:\pt{P^{\phantom{2}}_{\zeta}\,\rho\,P^{\phantom{2}}_{\zeta}}\quad.
\end{equation}
Assuming that $\rho\in{\cal M}_\zeta$ means, according to
definition~\eqref{M_zeta}, that it exists a quantum state $x\in{\cal
M}_N$ such that $\rho = P_{\zeta}\,x\,P_{\zeta}$\ , so that
\begin{equation}
\Tr\:\pq{\Phi_k\pt{\rho}}=\Tr\:\pt{P^2_{\zeta}\,x\,P^2_{\zeta}}=
\Tr\:\pt{P^{\phantom{2}}_{\zeta}\,x\,P^{\phantom{2}}_{\zeta}}=
\Tr\:\pt{\rho}\label{M_zeta_2}\quad,
\end{equation}
and $\Phi_k$ shows trace preservation on ${\cal M}_\zeta$. In
particular, for $k=N$, ${\cal M}_\zeta={\cal M}_N$
and~\eqref{M_zeta_2} holds in full generality.

\noindent$\mathbf{2)}$\quad  From the same lines of previous point
$\mathbf{(1)}$ one gets: $\displaystyle
\Tr\:\pq{\Phi\pt{\rho_\star}}=\Tr\:\pt{N\,\check{\sigma}_{\zeta}\,\mathds{I}_N/N}=\Tr\:\pt{P_{\zeta}}/N=k/N$.

\noindent$\mathbf{3)}$\quad We deduce from Lemma~\ref{Lemma5} that
\begin{equation}
\Phi\pt{\rho_\star} =\frac{1}{N}\sum_{\mu=1}^S
A_\mu^{\phantom{\dagger}}\:A_\mu^{\dagger}\quad\label{ks_tnimstar},
\end{equation}
that is positive and Hermitian too. Its normalization, computed in
previous point $\mathbf{(2)}$, makes this state a sub--normalized
state. The second statement in $\mathbf{(3)}$ is actually the
definition of \co{CP}
unital maps.\hfill$\qqeedd$ \\[2ex]
As a particular example, we will consider now the case of
$k$-extremal \co{CP}--\co{TNI} maps $\Phi_k$ associated with rescaled
dynamical matrices that are \emph{product states},
$\sigma_{\zeta}=\omega\otimes\check{\sigma}_{\zeta}$ for some
$\omega\in{\cal M}_N$ and $\zeta\in\mathcal{P}_k\pt{\Omega_N}$. For
this very special case, the action of $\Phi_k$ on a generic
$\rho\in{\cal M}_N$ reads
\begin{equation}
\Phi\pt{\rho}
=\omega\:\Tr\:\pt{P_{\zeta}\,\rho}\quad\label{prod_stat}:
\end{equation}
the map $\Phi_k$ sends the entire $k$--dimensional subspace spanned
by ${\cal M}_{\zeta}$ into $\omega$, whereas the complementary
subspace is annihilated.
\section{Concluding Remarks}%

In this work we investigated the set of subnormalized quantum states
and the set of completely positive, trace non--increasing maps. In
particular we computed the volume of the set of subnormalized states
and provided an algorithm to generate such states randomly with
respect to the \co{H}ilbert--\co{S}chmidt (Euclidean) measure.

We described the structure of the set of \co{CP--TNI} maps and
computed its volume. It is worth to emphasize that up till now no
exact result for the Hilbert-Schmidt volume of the  set of
\co{CP--TP} maps is known, (although some estimates can be done
\cite{Sza07}). On  the other hand, in this work we obtained an exact
result for the \co{HS} volume of the set of \co{TNI} maps, which
includes the set of \co{TP} maps.

Our paper contains several side results worth mentioning. In
particular, we found:\\[-3ex]
\begin{Ventry}{\ \ a)}
\item[\ \ a)] the probability distribution of states obtained by partial trace
              of mixed states of the bi--partite system distributed according to
              the \co{H}ilbert--\co{S}chmidt measure;\\[-3ex]
\item[\ \ b)] the volume of the sets of normalized, subnormalized
              and subtracial states with respect to the measures induced
              by partial trace;\\[-3ex]
\item[\ \ c)] an interpretation of the extremal trace non--increasing
              maps, which act as trace preserving on certain subspaces.
\end{Ventry}
\section*{Acknowledgments}%
It is a pleasure to thank  L. Hardy and S. Szarek for fruitful
discussions. We gratefully acknowledge financial support by the
\co{SFB}/Transregio-12 project financed by \co{DFG}, a grant number
1\, \co{P}03\co{B}\, 042\, 26 of the Polish Ministry of Science and
Information Technology and the European Research project \co{COCOS}.
\appendix
\section{Proof of Lemma~\ref{Lemma1}}%
\label{app_A1}
For any given $t\in\mathds{R}^+$, consider the quantity
\begin{align}
G\pt{t,N,K} &\coleq \int \Theta\pt{t-\sum_{i=1}^N
\Lambda_i}\prod_{i=1}^N \Theta(\Lambda_i)\:\Lambda_i^{K-N}
\prod_{i<j}
\pt{\Lambda_i-\Lambda_j}^2\;\ud\Lambda_1\,\ldots\,\ud\Lambda_N\quad,\label{lemma1_01}
\intertext{where $\Theta$ is the Heaviside step function, fulfilling
$\Theta\pt{tx}=\Theta\pt{x}$ for every positive $t$. By
 rescaling the variables $\Lambda_i\rightarrow t\lambda_i$\ we get}%
G\pt{t,N,K} &= t^{NK}\int \Theta\pt{1-\sum_{i=1}^N
\lambda_i}\prod_{i=1}^N \Theta(\lambda_i)\:\lambda_i^{K-N}
\prod_{i<j}
\pt{\lambda_i-\lambda_j}^2\;\ud\lambda_1\,\ldots\,\ud\lambda_N\quad\cdot\nonumber
\intertext{Comparing with~\eqref{lemma1_01} gives us the scaling relation
$\displaystyle G\pt{t,N,K} =t^{NK}\:G\pt{1,N,K}$. \ By taking the derivative at $t=1$, we obtain}%
 \left.\pq{\frac{\ud}{\ud t}\;G\pt{t,N,K}}\right|_{t=1} &=NK\cdot
G\pt{1,N,K} \label{lemma1_02}\quad,
\end{align}
and equation~\eqref{lemma1_01}, together with~\eqref{constab2},
yield
\begin{equation}
\left.\pq{\frac{\ud}{\ud t}\;G\pt{t,N,K}}\right|_{t=1} =\int
\delta\pt{1-\sum_{i=1}^N \Lambda_i}\prod_{i=1}^N
\Theta(\Lambda_i)\:\Lambda_i^{K-N} \prod_{i<j}
\pt{\Lambda_i-\Lambda_j}^2\;\ud\Lambda_1\,\ldots\,\ud\Lambda_N=\frac{1}{{C}_N^{(K-N+1,2)}
}\quad\cdot\nonumber
\end{equation}
Observe that $G\pt{1,N,K}$ on the r.h.s.~of~\eqref{lemma1_02} is
nothing but the integral in the statement of
Lemma~\ref{Lemma1}, so the result follows.\ \hfill$\qqeedd$ \\[2ex]
\begin{quote}
\begin{NNN}[Volume of $\bs{{\cal
M}_N^{\phantom{\text{\co{s}}}}}$ and $\bs{{\cal
M}_N^{\text{\co{sub}}}}$ with respect to induced measures]{}\ \\[-5.5ex]
\begin{Ventry}{htyr}\label{rem1}
    \item[]
The one parameter measure $\ud\nu_K(\Lambda_1,\ldots,\Lambda_N)$
introduced in Lemma~\ref{Lemma1} is called induced
measure~\cite{ZS01}, and appears as natural measure for reduced
density matrices of pure states equidistributed on a bipartite
system of dimension $N\times K$. As it is evident by
comparing~\eqref{dimu} and~\eqref{nu_K}, the \co{HS}--distribution
coincides with $\ud\nu_N(\Lambda_1,\ldots,\Lambda_N)$, and with
respect to this specific measure we compute the volumes of ${\cal
M}_N^{\phantom{\text{\co{s}}}}$ and of ${\cal
M}_N^{\text{\co{sub}}}$, given by equation~\eqref{volqs}
and~\eqref{volsub}. As a by--product of Lemma~\ref{Lemma1}, we are
able to compute the volumes of quantum states and that of
sub--normalized states also in the case of induced measure for
arbitrary $K$,
\begin{alignat}{2}
\mathrm{Vol}_{N,K}\pt{{\cal M}_N}&
=\frac{\sqrt{N}}{N\,!}\:\frac{\mathrm{Vol}\pq{Fl^{(N)}_{\mathds{C}}}}{C_N^{(K-N+1,2)}}&&=\sqrt{N}\:{(2\pi)}^{N(N-1)/2}
\:\frac{\Gamma\pt{K-N+1}\Gamma\pt{K-N+2}\cdots\Gamma\pt{K-1}\Gamma\pt{K}}{\Gamma\pt{N^2}}\quad\nonumber
\intertext{and} \mathrm{Vol}_{N,K}\pt{{\cal M}_N^{\text{\co{sub}}}}&
=\frac{1}{N\,!}\:\frac{\mathrm{Vol}\pq{Fl^{(N)}_{\mathds{C}}}}{NK\:C_N^{(K-N+1,2)}}&&=\phantom{\sqrt{N}}\:{(2\pi)}^{N(N-1)/2}
\:\frac{\Gamma\pt{K-N+1}\Gamma\pt{K-N+2}\cdots\Gamma\pt{K-1}\Gamma\pt{K}}{NK\;\Gamma\pt{N^2}}\quad\cdot\nonumber
\end{alignat}
\end{Ventry}
\end{NNN}
\end{quote}
\section{Proof of Lemma~\ref{Lemma2}}%
\label{app_A2}
Let us define $\varepsilon_{n\nu}^{(\Phi)}\coleq\Phi_{\supop{m
m}{n\nu}}=D_{\supop{m n}{m\nu}}$. From~\eqref{D_a} it follows that
$\varepsilon^{(\Phi)}$ is Hermitian.
For any given $\rho\in{\cal M}_N$, the hypothesis $\Phi\in{\cal
CP}^{\text{\co{TNI}}}_N$ implies $\rho_{\supop{mm}{}}^{\prime}
=\Phi_{\supop{mm}{n\nu}}^{\phantom{\prime}}
\rho_{\supop{n\nu}{}}^{\phantom{\prime}}=
\varepsilon_{\supop{n\nu}{}}^{(\Phi)}
\rho_{\supop{n\nu}{}}^{\phantom{\prime}}=\Tr\pt{\varepsilon^{(\Phi)}\rho}\leq
1=\Tr\pt{\mathds{I}_N\rho}$.\\
In other words
\begin{equation}
\Tr\pq{\pt{\mathds{I}_N-\varepsilon^{(\Phi)}}\rho}\geq0
\label{lemma2_1}\quad,\quad\forall\rho\in{\cal M}_N\quad\cdot%
\end{equation}
The matrix $\pt{\mathds{I}_N-\varepsilon^{(\Phi)}}$ is Hermitian,
thus there exists a unitary $U_{(\Phi)}^{\phantom{\dagger}}$ such that
$U_{(\Phi)}^{\phantom{\dagger}}\pt{\mathds{I}_N-\varepsilon^{(\Phi)}}U_{(\Phi)}^{\dagger}
=\Xi^{(\Phi)}$,
with $\Xi^{(\Phi)}$ diagonal. Equation~\eqref{lemma2_1} must hold
for all $\rho\in{\cal M}_N$, so that will hold in particular on the
sequence of density matrices
\begin{equation}
\rho^{(\Phi,i)}=U_{(\Phi)}^{\dagger}\ket{i}\bra{i}
U_{(\Phi)}^{\phantom{\dagger}}
\label{lemma2_2}\quad\cdot%
\end{equation}
Inserting each of the $N$ matrices of~\eqref{lemma2_2} into
equation~\eqref{lemma2_1}, and using the cyclic property of the trace, one
obtains
\begin{alignat}{2}
\Xi^{(\Phi)}_{ii}&\geq0 &\quad&,\quad\forall\ 1\leq
i\leq N\quad,\label{xi_ii}%
\intertext{or equivalently}%
 \varepsilon^{(\Phi)}_{ii}&\leq 1 &&,\quad\forall\ 1\leq
i\leq N\quad.\label{epsilon_ii}%
\end{alignat}
Now~\eqref{D_l2} follows by observing that
$\varepsilon^{(\Phi)}=N\:\Tr_A\,\sigma_{\Phi}$.\hfill$\qqeedd$ \\[2ex]
\section{Proof of Lemma~\ref{Lemma3}}%
\label{app_A3}
\noindent Directly from~(\ref{nu_Kcubo}--\ref{HJ5}) we write
\begin{align}
\int_{\Box_N}\ud\nu_K(\Lambda_1,\ldots,\Lambda_N) & =
\int\prod_{i=1}^N \Theta\pt{\frac{1}{N}-
\Lambda_i}\:\Theta(\Lambda_i)\:\Lambda_i^{K-N}\prod_{i<j}
\pt{\Lambda_i-\Lambda_j}^2\;\ud\Lambda_1\,\ldots\,\ud\Lambda_N \quad
\cdot\nonumber%
\intertext{By
 rescaling variables $\Lambda_i\rightarrow \lambda_i/N$\ , one obtains the result}%
\int\ud\nu_K(\Lambda_1,\ldots,\Lambda_N) &= N^{-NK}\int\prod_{i=1}^N
\Theta\pt{1-
\lambda_i}\:\Theta(\lambda_i)\:\lambda_i^{K-N}\prod_{i<j}
\pt{\lambda_i-\lambda_j}^2\;\ud\lambda_1\,\ldots\,\ud\lambda_N
\nonumber\\%
&= \frac{1}{N^{KN}}\int_0^1\cdots\int_0^1%
 \prod_{i=1}^N\lambda_i^{K-N}\prod_{i<j}
\pt{\lambda_i-\lambda_j}^2\;\ud\lambda_1\,\ldots\,\ud\lambda_N
\nonumber
\end{align}
equivalent to eq.~\eqref{HJ5} (see~\cite{Me91} for the definition of
the Selberg's integral).\hfill$\qqeedd$ \\[2ex]
\section{Proof of Lemma~\ref{Lemma4}}%
\label{app_A4}
\noindent In Section~\ref{tnim} we decided to express every
\co{CP}--\co{TNI} map $\Phi:\rho\longmapsto\rho^{\prime}$ in terms
of a linear superoperator $\Phi$ and a related dynamical matrix
$D_\Phi$, as follows
\begin{equation}
\rho^{\prime}_{\supop{ik}{}}=
\Phi^{\phantom{\prime}}_{\supop{ik}{j\ell}}\rho^{\phantom{\prime}}_{\supop{j\ell}{}}=
D^{\phantom{\prime}}_{\supop{ij}{k\ell}}\rho^{\phantom{\prime}}_{\supop{j\ell}{}}\label{L4_1}\quad.
\end{equation}
According to Lemma~\ref{Lemma2}, the $N^2\times N^2$ dynamical
matrix $D_\Phi$ is Hermitian and positive, thus one can benefit of
its spectral decomposition. Defining as
$\ket{m^{(\mu)}}\coleq\sum_{ij}^{\phantom{(\mu)}}m^{(\mu)}_{ij}\ket{i;j}$
the $\mu^{\text{th}}$ (bi--indexed) eigenvector of $D_\Phi$,
corresponding to the eigenvalue $m^{(\mu)}\in \mathds{R}^{+}$, the
square root decomposition reads
\begin{equation}
D_\Phi=\sum_{\mu=1}^{S}m^{(\mu)}\ket{m^{(\mu)}}\bra{m^{(\mu)}}=%
\sum_{\mu=1}^{S}\pt{\sqrt{m^{(\mu)}}\ket{m^{(\mu)}}}\pt{\sqrt{m^{(\mu)}}\bra{m^{(\mu)}}}=%
\sum_{\mu=1}^{S}\ket{A_\mu}\bra{A_\mu}\quad,\label{L4_2}
\end{equation}
with ${\pq{A_\mu}}_{ij}\coleq\sqrt{m^{(\mu)}}\;m^{(\mu)}_{ij}$ and
$S$ is given by the number of eigenvalues of $D_\Phi$ different from
zero. Equation~\eqref{L4_2} can be re--expressed in matrix form as
\begin{equation}
D_{\supop{ij}{k\ell}}=
 \sum_{\mu=1}^{S}{\pq{A_\mu}}_{ij}\overline{{\pq{A_\mu}}}_{k\ell}
\quad,\label{L4_3}
\end{equation}
where the symbol $\overline{x}$ denotes the complex conjugate of $x$.
Now we observe that the each bi--indexed vector ${\pq{A_\mu}}_{ij}$\
, of size $N^2$, can be seen as a square matrix of size $N$, and
reshuffling its indexes one gets
$\overline{{\pq{A_\mu^{\phantom{\dagger}}}}}_{k\ell}={\pq{A_\mu^{\dagger}}}_{\ell
k}$.\ Equation~\eqref{L4_1} becomes
\begin{equation}%
\rho^{\prime}_{\supop{ik}{}}=
D^{\phantom{\prime}}_{\supop{ij}{k\ell}}\rho^{\phantom{\prime}}_{\supop{j\ell}{}}%
=\sum_{\mu=1}^{S}{\pq{A_\mu}}_{ij}\;%
{\rho_{\phantom{j}}^{\phantom{j}}}_{\!\!\!j\ell}\:%
{\pq{{A_\mu^{\dagger}}}}_{\ell k} \label{L4_4}\quad,
\end{equation}
that is the matrix form of~\eqref{ks_tnim1}.

\noindent The explicit matrix form of the l.h.s. of~\eqref{ks_tnim2}
reads
\begin{equation}
 \sum_{\mu=1}^S{\pq{A_\mu^{\dagger}\:A_\mu^{\phantom{\dagger}}}}_{ij}=%
 \sum_{\mu=1}^S\sum_{\ell=1}^N{\pq{A_\mu^{\dagger}}}_{i\ell}{\pq{A_\mu^{\phantom{\dagger}}}}_{\ell
 j}=
 \sum_{\ell=1}^N\sum_{\mu=1}^S{\pq{A_\mu}}_{\ell j}\overline{{\pq{A_\mu}}}_{\ell
 i}=
 \sum_{\ell=1}^N
D_{\supop{\ell j}{\ell i}} ={\Big[\Tr_A D_\Phi\Big]}_{ji}
\quad,\label{L4_5}%
\end{equation}
where we made use of~\eqref{L4_3}, so that~\eqref{ks_tnim2} follows
from the Proposition of pag.~\pageref{repr1}.

To prove the second part of the Lemma, one simply notes that the
r.h.s of~\eqref{L4_2} always defines a positive and Hermitian
$N^2\times N^2$ matrix $D$, no matter of which $S$ complex $N\times
N$ matrices have been used. The additional constraint that makes $D$
becoming a dynamical matrix, representative of some
\co{CP}--\co{TNI} map $\Phi$, is given by~\eqref{D_l2_fin}.
Imposing~\eqref{D_l2_fin}, in terms of the family of matrices
$A_\mu$, is equivalent of imposing~\eqref{ks_tnim_inv1}, as it can
be earned by reversing the chain~\eqref{L4_5} and by noting that
$\check{\sigma}\in{\cal
M}_N^{\Box}\Longrightarrow\check{\sigma}^{\text{\co{T}}}\in{\cal
M}_N^{\Box}$. Finally the claim can be obtained through the same
steps~(\ref{L4_2}--\ref{L4_4}) as before and this ends the proof.\hfill$\qqeedd$ \\[2ex]

\end{document}